\documentclass[12pt,aps,prd,preprint,tightenlines,superscriptaddress,
   showpacs,nofootinbib]{revtex4}
\newcommand{\PRE}[1]{{#1}} 

\usepackage{bm}
\usepackage{graphicx}

\graphicspath{{figs/}}

\newcommand{\gweak}{g_{\text{weak}}}
\newcommand{\mweak}{m_{\text{weak}}}

\newcommand{\sigmaSI}{\sigma_{\text{SI}}}

\newcommand{\ipb}{\text{pb}^{-1}}
\newcommand{\ifb}{\text{fb}^{-1}}

\newcommand{\kev}{\text{keV}}

\newcommand{\gev}{\text{GeV}}
\newcommand{\tev}{\text{TeV}}
\newcommand{\pb}{\text{pb}}

\newcommand{\eg}{{\em e.g.}}

\newcommand{\eqref}[1]{Eq.~(\ref{#1})}

\newcommand{\Eqref}[1]{Equation~(\ref{#1})}
\newcommand{\secref}[1]{Sec.~\ref{sec:#1}}

\newcommand{\figref}[1]{Fig.~\ref{fig:#1}}

\newcommand{\Figref}[1]{Figure~\ref{fig:#1}}
\newcommand{\tableref}[1]{Table~\ref{table:#1}}
\newcommand{\tablesref}[2]{Tables~\ref{table:#1} and \ref{table:#2}}
\newcommand{\Dsle}[1]{\slash\hskip -0.28 cm #1}
\newcommand{\met}{{\Dsle E_T}}
\newcommand{\Dslp}[1]{\slash\hskip -0.23 cm #1}
\newcommand{\mpt}{{\Dslp p_T}}

\newcommand{\mmess}{M_{\text{mess}}}

\newcommand{\be}{\begin{equation}}
\newcommand{\ee}{\end{equation}}
\newcommand{\bea}{\begin{eqnarray}}
\newcommand{\eea}{\end{eqnarray}}

\begin{document}

\preprint{UCI-TR-2009-14}
\preprint{UH-511-1146-10}

\title{
\PRE{\vspace*{0.8in}}
Dark Matter-Motivated Searches for Exotic 4th Generation Quarks in
Tevatron and Early LHC Data
\PRE{\vspace*{0.3in}}
}

\author{Johan Alwall}
\affiliation{Department of Physics and National Center for
Theoretical Sciences, National Taiwan University, Taipei, 10617, Taiwan
\PRE{\vspace*{.1in}}
}

\author{Jonathan L.~Feng}
\affiliation{Department of Physics and Astronomy, University of
California, Irvine, CA 92697, USA
\PRE{\vspace*{.1in}}
}

\author{Jason Kumar}
\affiliation{Department of Physics and Astronomy, University of
Hawai'i, Honolulu, HI 96822, USA
\PRE{\vspace*{.1in}}
}

\author{Shufang Su%
\PRE{\vspace*{.4in}}
}
\affiliation{Department of Physics, University of Arizona, Tucson, AZ
85721, USA
\PRE{\vspace*{.5in}}
}

\date{February 2010}

\begin{abstract}
\PRE{\vspace*{.3in}} We determine the prospects for finding dark
matter at the Tevatron and LHC through the production of exotic 4th
generation quarks $T'$ that decay through $T' \to t X$, where $X$ is
dark matter.  The resulting signal of $t \bar{t} + \met$ has not
previously been considered in searches for 4th generation quarks, but
there are both general and specific dark matter motivations for this
signal, and with slight modifications, this analysis applies to any
scenario where invisible particles are produced in association with
top quarks.  Current direct and indirect bounds on such exotic quarks
restrict their masses to be between 300 and 600 GeV, and the dark
matter's mass may be anywhere below $m_{T'}$.  We simulate the signal
and main backgrounds with MadGraph/MadEvent-Pythia-PGS4.  For the
Tevatron, we find that an integrated luminosity of $20~\ifb$ will
allow 3$\sigma$ discovery up to $m_{T'} = 400~\gev$ and 95\% exclusion
up to $m_{T'} = 455~\gev$.  For the 10 TeV LHC with $300~\ipb$, the
discovery and exclusion sensitivities rise to 490 GeV and 600 GeV.
These scenarios are therefore among the most promising for dark matter
at colliders.  Perhaps most interestingly, we find that dark matter
models that can explain results from the DAMA, CDMS and CoGeNT
Collaborations can be tested with high statistical significance using
data already collected at the Tevatron and have extraordinarily
promising implications for early runs of the LHC.
\end{abstract}

\maketitle

\section{Introduction}

One of the great hopes for current and future particle colliders is
that they will be able to produce dark matter.  In this study, we
determine the prospects for finding dark matter through the production
of exotic 4th generation quarks $T'$ that decay through $T' \to t X$,
where $X$ is dark matter.  Current direct and indirect bounds on such
exotic quarks restrict their mass range to $300~\gev \alt m_{T'}\alt
600~\gev$.  Our analysis is valid for all dark matter masses up to
$m_{T'}-m_W-m_b$, although there are special reasons to be interested
in very light $X$ particles, with $m_X \sim 1 - 10~\gev$.

There are both general and specific dark matter motivations for this
signal.  Starting with the general motivation, one of the few things
that is absolutely certain about dark matter is that it must be
long-lived on cosmological time scales.  This is typically achieved by
giving dark matter a charge under an unbroken discrete or continuous
symmetry, which makes it absolutely stable.  None of the unbroken
symmetries of the standard model (SM) will do for this purpose, so the
dark matter particle must be charged under a new unbroken symmetry.

There are then two options.  The dark matter, with its stabilizing
``dark charge,'' may have only gravitational interactions with the SM.
In this case, dark matter may have interesting astrophysical
signals~\cite{Ackerman:2008gi,Feng:2009mn}, but it cannot be
discovered at colliders.

Alternatively, the dark matter may be coupled to SM particles $f$
through connector particles $Y$ that have both dark and SM charges to
make $XYf$ couplings possible.  $X$ may or may not have SM weak
interactions.  However, $Y$ necessarily has SM charge.  It can
therefore be produced at colliders, and so dark matter can be
discovered through $Y$ production followed by $Y \to fX$. Since the
energy frontier is dominated by hadron colliders for the foreseeable
future, the most promising case is where $Y$ is strongly interacting.
Supersymmetric (universal extra dimension) models provide a concrete
example of this, where the dark matter is
neutralinos~\cite{Goldberg:1983nd,Ellis:1983ew} (Kaluza-Klein (KK)
gauge bosons~\cite{Servant:2002aq,Cheng:2002ej}), the connector
particles are squarks (KK quarks), and the stabilizing symmetry is
$R$-parity (KK-parity). Here we consider the case where the dark
matter has no SM gauge interactions, the connector particles are
exotic quarks, and the stabilizing symmetry may be either discrete or
continuous~\cite{Feng:2008ya,Feng:2008dz}.  Note, however, that with
minor modifications, our analysis applies much more generally, both to
the supersymmetric and extra dimensional scenarios just mentioned, as
well as to many other dark matter scenarios motivated by the general
chain of reasoning given above.

These scenarios also have a more speculative, but at the same time
more specific and tantalizing, dark matter motivation. The DAMA
experiment sees an 8.9$\sigma$ signal in the annual modulation of
scattering rates that can potentially be explained by dark
matter~\cite{Bernabei:2008yi,Bernabei:2010mq}.  Uncertainties from
both astrophysics~\cite{astro} and detector response~\cite{channeling}
open the possibility that DAMA can be explained without conflicting
with other experiments by a light dark matter particle with mass $m_X
\sim 1-10~\gev$ elastically scattering off nucleons with
spin-independent cross section $\sigmaSI \sim 10^{-2} -
10^{-5}~\pb$~\cite{DAMAlowmass}.  This explanation is supported by
unexplained events recently reported by the CoGeNT
Collaboration~\cite{Aalseth:2010vx}, which, if interpreted as a dark
matter signal, are best fit by dark matter with $m_X \sim 9~\gev$ and
$\sigmaSI \sim 6.7 \times 10^{-5}~\pb$.  The low mass and high cross
sections preferred by DAMA and CoGeNT are consistent with the recent
bounds and results from CDMS~\cite{Ahmed:2009zw}.

Such large cross sections are several orders of magnitude larger than
those of typical weakly-interacting massive particles (WIMPs).  As we
review below, however, they are easily obtained if the dark matter
particle scatters through $X q \to Q' \to X q$, where $Q'$ is an
exotic 4th generation quark.  Furthermore, this scenario naturally
emerges in WIMPless dark matter models, where dark matter not only can
have the correct $m_X$ and $\sigmaSI$ for DAMA and CoGeNT, but also
naturally has the correct thermal relic
density~\cite{Feng:2008ya,Feng:2008dz}.  This scenario is a special
case of the general framework described above, and, as discussed in
\secref{model}, particularly motivates the case where dark matter
couples to third generation quarks and the $X$ particles are light.

Given all of these motivations, we explore here the detection
prospects for exotic 4th generation quarks decaying directly to dark
matter.  This signal differs from most supersymmetry searches, which
typically assume that decays to dark matter are dominated by cascade
decays.  The 4th generation quarks examined here also differ from the
4th generation quarks that are typically studied, because they are
charged under a new symmetry under which SM particles are neutral.
This forbids decays to SM quarks, such as $T' \to Wb$ and $B' \to Wt$,
which are the basis for most standard 4th generation quark searches.
Instead, all $Q'$ decays must necessarily produce hidden sector
particles charged under the new symmetry, with the lightest such
particle being $X$, the dark matter.  This leads to the signal of
$t\bar{t} + \met$, which has previously not been considered in
searches for 4th generation quarks.  In addition, as emphasized
earlier, this type of signals appear in a general set of dark matter
motivated models, as well as other new physics scenarios, such as
little Higgs models with $T$-parity conservation~\cite{littleHiggsT}
and models in which baryon and lepton number are gauge
symmetries~\cite{FileviezPerez:2010gw}.  The results of our analyses
can be easily applied to these other models.  We will find that $T'$
pair production followed by $T' \to tX$ leads to $\met$ signals that
may be discovered at the Tevatron with $10~\ifb$ of integrated
luminosity, or at the LHC with integrated luminosities of as little as
$\sim 100~\ipb$.

In \secref{model} we detail the dark matter motivations and define the
model we explore.  In \secref{constraints} we summarize the current
theoretical and experimental constraints on exotic 4th generation
quarks.  We then describe the details of our signal and background
simulations and cut analysis in \secref{simulation}.  In
\secref{discovery} we present the prospects for discovering or
excluding these exotic 4th generation quark scenarios with Tevatron
and early LHC data.  We conclude with a discussion of future prospects
in \secref{conclusions}.

\section{Dark Matter Models with Exotic Quarks}
\label{sec:model}

\subsection{WIMPless Dark Matter}

As discussed above, in this work we consider the case where dark
matter has a charge under some new symmetry, but no SM gauge
interactions. This dark matter is therefore not a typical WIMP, but it
may nevertheless naturally appear in theories motivated by the gauge
hierarchy problem and have the correct thermal relic density. This is
the case for WIMPless dark matter models~\cite{Feng:2008ya},
supersymmetric models where the effects of a supersymmetry-breaking
sector are transmitted to both the minimal supersymmetric SM (MSSM)
sector and a hidden sector through gauge-mediation.  The hidden
sector superpartner mass scale is
\begin{equation}
m_{\text{hidden}} \propto g_{\text{hidden}}^2 {F \over \mmess} \ ,
\end{equation}
where $g_{\text{hidden}}$ is the hidden sector gauge coupling, $F$ is
the supersymmetry-breaking scale squared, and $\mmess$ is the mass
scale of the messenger particles.  Because the MSSM superpartner
masses are also generated by gauge-mediation from the same
SUSY-breaking sector, we find
\begin{equation}
{g_{\text{hidden}}^2 \over m_{\text{hidden}}}
\sim {\gweak^2 \over \mweak} \sim {\mmess\over F}  \ .
\label{thermalsigmascaling}
\end{equation}
The ratio $g^4 / m^2$ sets the annihilation cross section of a
particle through gauge interactions, which in turn determines the
thermal relic density of a stable particle~\cite{Zeldovich:1965}.  The
``WIMP miracle'' is the remarkable coincidence that, for a stable WIMP
with mass $m \sim \mweak$ and coupling $g \sim \gweak$, this thermal
relic density is roughly that required by astrophysical observations.
\Eqref{thermalsigmascaling} shows that our hidden sector candidate
automatically has approximately the same annihilation cross section,
and thus the same relic density.

If there are connectors $Y$ with both dark and SM charge, WIMPless
dark matter may have observable interactions through couplings $XYf$,
where $f$ are SM particles.  In this case, WIMPless dark matter has
many of the virtues and implications commonly associated with WIMPs.
In contrast to WIMPs, however, the WIMPless dark matter's mass need
not be at the electroweak symmetry-breaking scale.  It may be treated
as a free parameter, which in turn determines the gauge coupling
strength of the hidden sector.  This freedom opens new possibilities
for dark matter model parameters and new experimental search windows.

\subsection{Explaining DAMA}

The DAMA dark matter signal of annual modulation in direct detection
has motivated a variety of explanations~\cite{damamodels}.  The
canonical possibility, where WIMP dark matter with mass $\sim
100~\gev$ elastically scatters, is excluded, as the required
scattering cross section is in conflict with other experiments.
Inelastic scattering~\cite{Han:1997wn,Hall:1997ah,Smith:2001hy}, in
which dark matter is assumed to scatter to another state that is $\sim
100~\kev$ heavier, has also been explored.  Such scattering alleviates
the conflict between different direct detection experiments, but is
tightly constrained by neutrino bounds on dark matter annihilation in
the sun~\cite{Nussinov:2009ft,Menon:2009qj}.

An alternative explanation is elastic scattering of a light dark
matter particle with $m \sim 1-10~\gev$ and large spin-independent
nucleon scattering cross section $\sigmaSI \sim 10^{-2} -
10^{-5}~\pb$~\cite{DAMAlowmass}, a region also supported by recent
results from CoGeNT~\cite{Aalseth:2010vx}.  Such explanations are
possible if the DAMA signal is enhanced by populations of dark matter
in tidal streams~\cite{astro} or detection thresholds are lowered by
channeling~\cite{channeling}, and may also require an unusual
background spectrum for consistency~\cite{Kudryavtsev:2009gd}.  In
typical WIMP models, the required small masses and large cross
sections are possible~\cite{Bottino:2003iu}, but not at all generic.

In contrast, WIMPless dark matter provides a natural setting for the
low mass explanation.  In the example presented in
Refs.~\cite{Feng:2008ya,Feng:2008dz}, the dark matter particle in the
hidden sector couples to the SM through Yukawa couplings
\begin{equation}
V = \lambda \left[ X \bar{Q}'_L q_L + X \bar{B}'_R b_R
+ X \bar{T}'_R t_R \right] \ .
\label{couplings}
\end{equation}
Each term may have a different coupling, but we assume equal couplings
for simplicity.  In \eqref{couplings}, $X$ is the dark matter, a
complex scalar\footnote{In a non-supersymmetric context where the
stabilizing symmetry is discrete, $X$ could also be a real scalar.}
charged under a discrete symmetry (hidden parity); $q_L^T \equiv (t_L,
b_L)$, $t_R$, and $b_R$ are the third generation quarks of the SM; and
${Q'}_L^T \equiv (T'_L, B'_L)$, $T'_R$, and $B'_R$ are the connectors,
exotic 4th generation quarks.  The $Q'$ have hidden parity and are in
the SM SU(3)$\times$SU(2)$\times$U(1)$_Y$ representations
\begin{eqnarray}
Q'_L &:& \left( 3 ,2, {\textstyle \frac{1}{6}} \right) \nonumber\\
T'_R &:& \left( 3, 1, {\textstyle \frac{2}{3}} \right) \nonumber\\
B'_R &:& \left( 3, 1, - {\textstyle \frac{1}{3}} \right) \ .
\end{eqnarray}
The subscripts $L$ and $R$ refer to SU(2) doublets and singlets,
respectively, not chirality; the chirality of the $Q'_L$, $T'_R$, and
$B'_R$ fields is opposite to their SM counterparts, and they are
therefore mirror quarks.  Finally, the $Q'$ receive mass through
electroweak symmetry breaking.

The couplings of \eqref{couplings} imply scattering through $X q \to
Q' \to X q$, where $q = b, t$.  This induces a coupling to the gluons
of the nucleon at one-loop~\cite{Shifman:1978zn}.  As shown in
Ref.~\cite{Feng:2008dz}, for $m_X \sim 1 - 10~\gev$, $m_{Q'} \sim
300-500~\gev$, and $\lambda \sim 0.3-1$, the coupling to $b$ quarks
produces a cross section $\sigmaSI$ in the right range to explain DAMA
and CoGeNT.  For example, the best fit point for the CoGeNT data can
be obtained with $m_X \sim 9~\gev$, $m_{Q'} \sim 400~\gev$ and
$\lambda \sim 0.7$.  The large cross section can be understood as
follows: spin-independent scattering requires a chirality flip on the
fermion line, which is typically suppressed by a small Yukawa
coupling.  But if the dark matter is a scalar, than a mass insertion
on the $Q'$ propagator provides the necessary chirality flip without
suppressing $\sigmaSI$, since the $Q'$ are heavy and their Yukawa
couplings are large. The explanation therefore requires that $X$ is a
scalar and the connectors are chiral fermions.

In fact, this mechanism is overly efficient.  If the 3rd generation
quarks are replaced by 1st generation quarks in \eqref{couplings}, the
dark matter couples to nucleons at tree-level, and the desired
$\sigmaSI$ is achieved for couplings $\lambda \sim
0.03$~\cite{Feng:2008ya}.  This is also perfectly acceptable and worth
studying~\cite{fermdm}, but the required coupling for $b$ quarks
appears to be somewhat more natural and, in the general case where
there are couplings to more than one generation, less constrained by
flavor-changing neutral currents.  In this study, we assume negligible
couplings to 1st and 2nd generation quarks and focus on the collider
phenomenology of the case where the $Q'$ decay directly to 3rd
generation quarks.

If the dark matter is stabilized not by a discrete symmetry, but by a
continuous symmetry, $Q'_L$ and $T'_R/B'_R$ must have opposite dark
charges to allow them to get a mass.  The Yukawa couplings of
\eqref{couplings} must therefore be generalized to
\begin{equation}
V' = \lambda \left[ X_L \bar{Q}'_L q_L + X_R \bar{B}'_R b_R
+ X_R \bar{T}'_R t_R \right] \ ,
\end{equation}
where $X_L$ and $X_R$ are two complex scalars
with opposite dark charges.  In general, $X_L$ and $X_R^*$
will mix to form mass eigenstates $X_1$ and $X_2$. The lighter state
is the dark matter particle and can couple to both $Q'_L$ and
$T'_R/B'_R$.  Despite slight additional complications, we therefore
recover the discrete symmetry case, although there may now be
additional decays $Q' \to q X_2$.  For simplicity, we focus in the
rest of this study on the discrete symmetry case with couplings given
in \eqref{couplings}.

\section{Existing Constraints}
\label{sec:constraints}

As with SM quarks, the 4th generation quarks receive their mass
through electroweak symmetry breaking, and so $m_{Q'} = y_{Q'} v
/ \sqrt{2}$, where $v \simeq 246~\gev$.  Perturbativity places an
upper bound on $m_{Q'}$; requiring $\alpha_{Q'} \equiv
y_{Q'}^2/{4\pi} \alt 1$ implies $m_{Q'} \alt 600~\gev$.  4th
generation quark masses are also constrained by precision electroweak
data.  These constraints are not modified by the exotic and mirror
features of the quarks we consider, and they imply $|m_{T'} - m_{B'}|
\sim 50~\gev$, where some non-degeneracy is
required~\cite{Kribs:2007nz}.  4th generation quarks may also have
many beneficial effects, for example, raising the Higgs boson mass in
supersymmetric theories through their loop corrections, and enhancing
Higgs boson production rates~\cite{Kribs:2007nz}.

Direct searches place lower bounds on $m_{Q'}$.  These searches are
rather independent of the details of the couplings to the hidden
sector.  $T'$ and $B'$ production is dominated by QCD processes.  In
addition, the coupling $\lambda$ only affects the $T'$ and $B'$
lifetimes.  For all but extremely small $\lambda$, the $T'$ and $B'$
decay promptly.

With the model framework and assumptions given in \secref{model}, the
possible decays of the $Q'$ are $T'\to t X$, $B'\to b X$, $T'\to
W^{+(*)} B'$, and $B'\to W^{-(*)} T'$.  If $|m_{T'} - m_{B'}| < m_W$,
the decays $T'\to W^{+ \, *} B'$ and $B'\to W^{-\, *} T'$ are strongly
suppressed by kinematics. In the type of scenarios we are interested
in here, the coupling between the $Q'$ and $X$ states is furthermore
rather strong, which means that the decays $T'\to t X$ and $B'\to b X$
completely dominate.  In our analysis below, we assume that $B(T' \to
t X) = B(B' \to b X) = 1$.

$B'$ pair production followed by $B'\to b X$ leads to a signature of
$2b + \met$, which is identical to the final state of bottom squark
pair production followed by $\tilde{b}\to b \tilde \chi_1^0$.
Searches for this supersymmetric signal have been carried at both CDF
and D\O\ at the Tevatron.  The D\O\ analysis, based on an integrated
luminosity of $310~\ipb$ from Run II, implies $m_{\tilde{b}} >
222~\gev$ (95\% CL) for $m_{\tilde \chi_1^0} < 50~\gev$~\cite{D0sb};
the corresponding CDF result using $295~\ipb$ is $m_{\tilde{b}} >
193~\gev$ (95\% CL)~\cite{CDFsb}.  Taking into account the difference
in $B' \bar{B'}$ and $\tilde{b} \tilde{b}^*$ cross sections, the D\O\
results imply $m_{B'} \agt 330~\gev$.

A later search for gluino pair production with $\tilde{g}\to b
\tilde{b}$ and $\tilde{b}\to b \tilde\chi_1^0$ has been carried out by
the CDF Collaboration using $2.5~\ifb$ collected luminosity.
Candidate events were selected requiring two or more jets, large
$\met$, and at least two $b$-tags~\cite{CDFgluino}.  Using neural net
analyses, CDF found $m_{\tilde{g}} > 350~\gev$ (95\% CL) for large
mass splitting $\Delta m = m_{\tilde{g}}-m_{\tilde{b}} \agt 80~\gev$,
and about 340 GeV for small $\Delta m \sim 20~\gev$.  Their result in
the case of small mass splitting $\Delta m$ can be applied to the
$B'\bar{B}'$ search, implying roughly $m_{B'} \agt 370~\gev$.
Finally, there are also projections for squark searches at the LHC
based on the $2j + \met$ signal~\cite{CMS08005}.
It is hard, however, to apply their results to the $2b +\met$ signal,
since no $b$-tagging is used in that analysis.

In our analysis, we will focus on $T'$ pair production, $ pp
(p\bar{p}) \to T'\bar{T'} \to t\bar{t}XX$, with the distinctive, but
more complicated, final state of a top quark pair plus missing energy
from the $X$ particles.  The CDF Collaboration has reported a search
for the analogous supersymmetric process of top squark pair production
based on an integrated luminosity of $2.7~\ifb$, using the purely
leptonic final states from $p\bar{p} \to \tilde{t}_1\tilde{t}_1^*$,
followed by $\tilde{t}_1\to b \tilde\chi_1^\pm \to b \tilde\chi_1^0 l
\nu$~\cite{CDFstop}.  The data are consistent with the SM, leading to
the constraint $m_{\tilde{t}_1} \agt 150 - 185~\gev$, where the exact
limit depends on $m_{\tilde\chi_1^0}$, $m_{\tilde\chi_1^\pm}$ and
$B(\tilde\chi_1^\pm\to \chi_1^0 l^\pm \nu)$.  Similar signals also
appear in other new physics scenarios such as little Higgs models with
$T$-parity~\cite{littleHiggsT}.  The $t\bar t + \met$ signature at the
LHC in the semi-leptonic channel has also been studied in
Ref.~\cite{HanGY}; however, that study focused on higher masses with
larger integrated luminosity. The hadronic mode has been analyzed at
the parton level in Ref.~\cite{MeadeDW}, focusing on the prospects for
spin determination and mass measurements.  In contrast, our study is
performed at the detector level and is focused on the exclusion and
discovery potential of both the Tevatron and early LHC data.

\section{Event Simulation, Backgrounds, and Cuts}
\label{sec:simulation}

\subsection{Simulation}

To investigate the discovery and exclusion prospects, we have
simulated production and decay of the new particles at the Tevatron
and at the LHC with $\sqrt{s} = 10~\tev$, as well as the main
backgrounds.  All simulations have been done using MadGraph/MadEvent -
Pythia 6.4.20 - PGS4~\cite{MadGraph,PYTHIA,pgs} with the $p_T$-ordered
Pythia showers and the CDF or ATLAS detector cards for PGS4. Matrix
element/parton shower matching has been applied both for signal and
backgrounds, and its validity has been double-checked by comparing
different maximum multiplicity samples. The parton distribution
function set used is CTEQ6L1, and factorization and renormalization
scales are set to $\mu_F^2 = \mu_R^2 = m_T^2 = m^2 + p_T^2$ for the
centrally-produced particle pair. We do not apply $K$-factors for
higher order QCD effects to either signal or background. These
$K$-factors are expected to enhance these cross sections and be
similar for the signal and top pair production (the dominant
background after cuts), and hence the effect of including them would
only be to increase the signal significance.  In addition, there are
uncertainties in the cross sections of both signal and backgrounds due
to parton distribution functions \cite{Martin:2002aw}.

For $m_{T'}-m_X < m_t$, the $T'$ cannot decay to an on-shell
$t+X$. These parameter points have therefore been simulated in
MadGraph/MadEvent using off-shell top decay, $ T'\bar{T'} \to b W^+X
\bar b W^- X$. This procedure guarantees that finite width effects are
correctly accounted for.

Note that QCD multi-jet backgrounds have not been simulated. Instead,
we refer to the studies of Refs.~\cite{QCDbgTeV,QCDbgLHC} and apply
similar cuts, in particular $\Delta\phi(\mpt, p_T^j)$ cuts, $\met$
cuts, and cuts on number of jets, which should be enough to suppress
the QCD multi-jet backgrounds to negligible levels.

The signatures of $pp (p\bar p) \to T'\bar{T'}$ are determined by the
decays of the top quark pair and can therefore be divided into
hadronic, semi-leptonic and purely leptonic channels. Since the $T'$
decays will always generate missing transverse momentum from the
invisible $X$ particles in the final state, the most relevant
backgrounds are those with significant missing energy from $W$ or $Z$
boson decays into neutrinos.

We will focus on the semi-leptonic and hadronic channels.  The
dilepton channel has suppressed cross section because of the small
leptonic decay branching ratios.  The semi-leptonic decay has the
advantage that the presence of an isolated lepton (electron or muon)
makes it easier to suppress QCD backgrounds, while for the fully
hadronic channel this requires additional $\Delta\phi(\mpt, p_T^j)$
cuts.  This is particularly important for early running of the LHC,
where the missing energy resolution might not yet be fully under
control. On the other hand, the hadronic decay mode has larger
branching ratio.  Moreover, the main backgrounds, $t\bar t$ and
$W^\pm$ production, only have substantial $\met$ in association with
leptons; in fact, the only truly irreducible background to the fully
hadronic decay mode is $Z\to\nu\bar\nu$ + jets (and the negligible
$t\bar t\nu\bar\nu$ background).

\subsection{Semi-leptonic Channel}

To distinguish signal from background in the semi-leptonic case, we
look for large $\met$ and large transverse mass of the leptonic $W$
candidate, defined to be $m_T^W \equiv m_T(p_T^l, \mpt) = \sqrt{2
|p_T^l| |\mpt| \cos(\Delta\phi(p_T^l, \mpt))}\ $.  Since the $\mpt$
for the background is mainly from $W$ decay, we expect $m_T^W < m_W$
for most of the background, while much of the signal extends beyond
this limit. For the background surviving the $\met$ and transverse
mass cuts, we expect a significant fraction to be due to decays of the
second top into hadronic $\tau$ leptons and missing energy. We
therefore expect fewer jets for the background than for the signal. To
further suppress the background, we also require the presence of a
second, hadronically-decaying $W$.

To implement this strategy, we employ the following precuts
(differences between the Tevatron and the LHC are noted where they
apply):
\begin{itemize}
\setlength{\itemsep}{1pt}\setlength{\parskip}{0pt}\setlength{\parsep}{0pt}
\item One isolated  electron or muon  with $|p_T^l| > 10~\gev$.
\item No additional isolated leptons with $|p_T^l| > 2~\gev$.
\item Minimum missing transverse energy: $\met > 100~\gev$.
\item Minimum transverse mass: $m_T^W > 100~\gev$.
\item At least 4 jets with $|p_T^j| > 20~\gev$ (Tevatron) or $|p_T^j|
> 40$ GeV (LHC).
\item At least one jet pair with invariant mass within the $W$ mass
window $|m_{jj} - m_W| < 10~\gev$.
\end{itemize}
We also use additional cuts to achieve the best signal significance:
\begin{itemize}
\setlength{\itemsep}{1pt}\setlength{\parskip}{0pt}\setlength{\parsep}{0pt}
\item Additional $m_T^W$ cut: $m_T^W > 150~\gev$
(Tevatron) or $m_T^W > 150, 200~\gev$ (LHC).
\item Additional $\met$ cuts: $\met > 150~\gev$ (Tevatron); $\met >
150,\;200,\;250~\gev$ (LHC).
\item $H_T = \sum_{i=1}^4 |p_T^j|_i + |p_T^l|$ cuts: $H_T > 300~\gev$
(Tevatron); $H_T > 400,\;500~\gev$ (LHC).
\item Combinations of the cuts above.
\end{itemize}

The relevant backgrounds for the semi-leptonic channel are $t\bar t$
(semi-leptonic and purely leptonic decays) and leptonically-decaying
$W^\pm$ + jets production. Top pairs are the main background.  Because
this has the same number of $b$-quarks as the signal, no $b$-tagging
information is employed, since this would suppress signal and
background by the same amount, leading to a reduced signal
significance.  For completeness, we also simulated $Z$ + jets (where
the largest contribution comes from $Z\to \tau^+\tau^-$ with one of
the taus decaying leptonically) and $t\bar t Z$, but these processes
both turned out to be negligible after precuts.

We show the distributions for missing transverse energy $\met$,
transverse mass $m_T^W$, number of jets $N(\text{jets})$,
and jet pair invariant mass $m_{jj}$ for example signal parameters and
backgrounds for the 10 TeV LHC in \figref{1lep-distributions}. Each
observable is plotted after the cuts coming before it in the list, and
the position of the precut is marked with a vertical dashed line. For
clarity, we have split the $t\bar t$ background into components:
semi-leptonic decays (to electron or muon), decays with at least one
tau lepton, purely leptonic decays (where both $W$'s decay to electron
or muon) and fully hadronic decay (which is negligible with these
cuts). The $m_{jj}$ plot shows the invariant mass for the jet pair
closest to the $W$ mass.  The corresponding distributions for the
Tevatron are qualitatively similar.

\begin{figure}[tb]
\begin{center}
\includegraphics*[width=0.48\columnwidth]{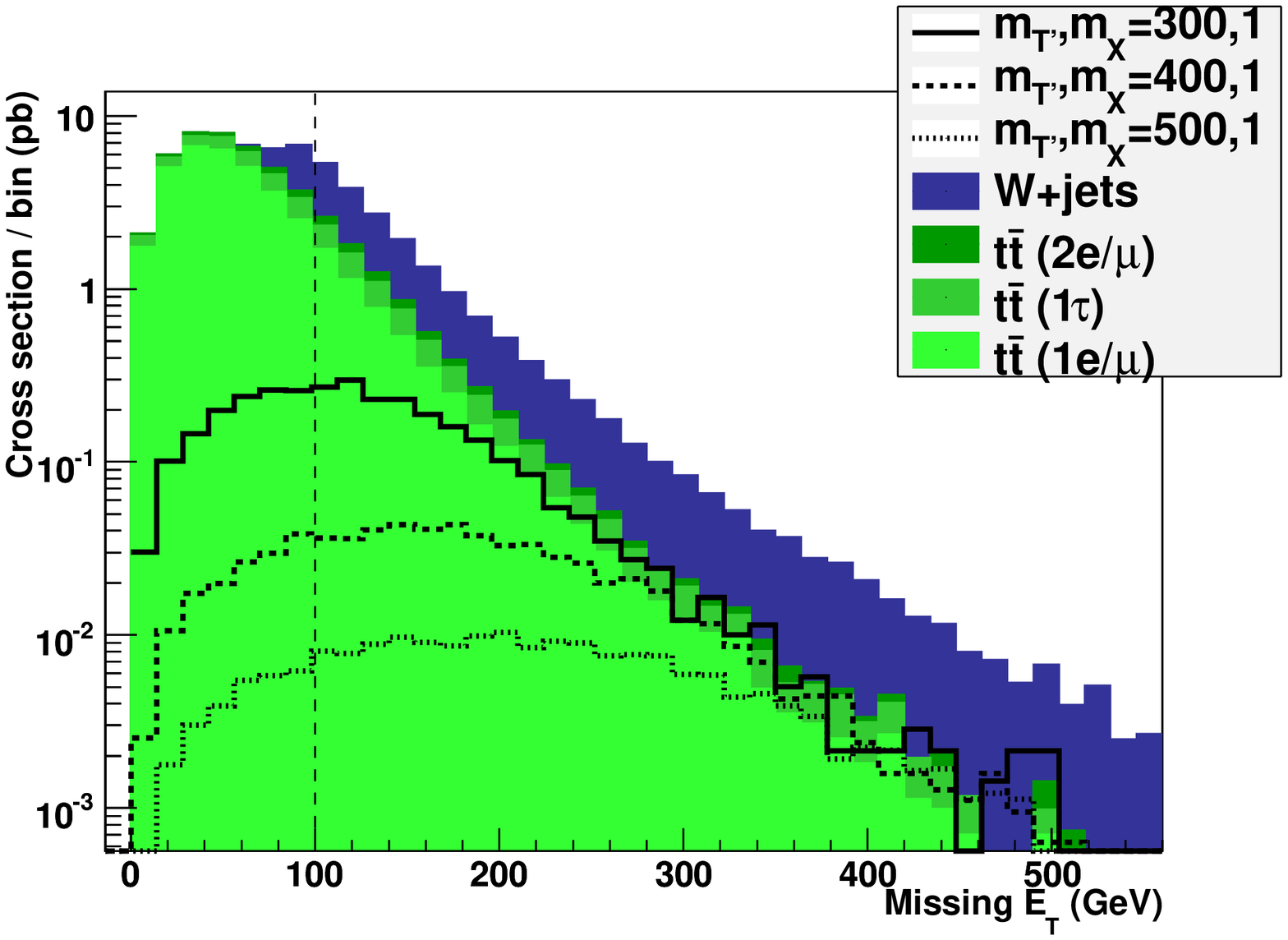}
\includegraphics*[width=0.48\columnwidth]{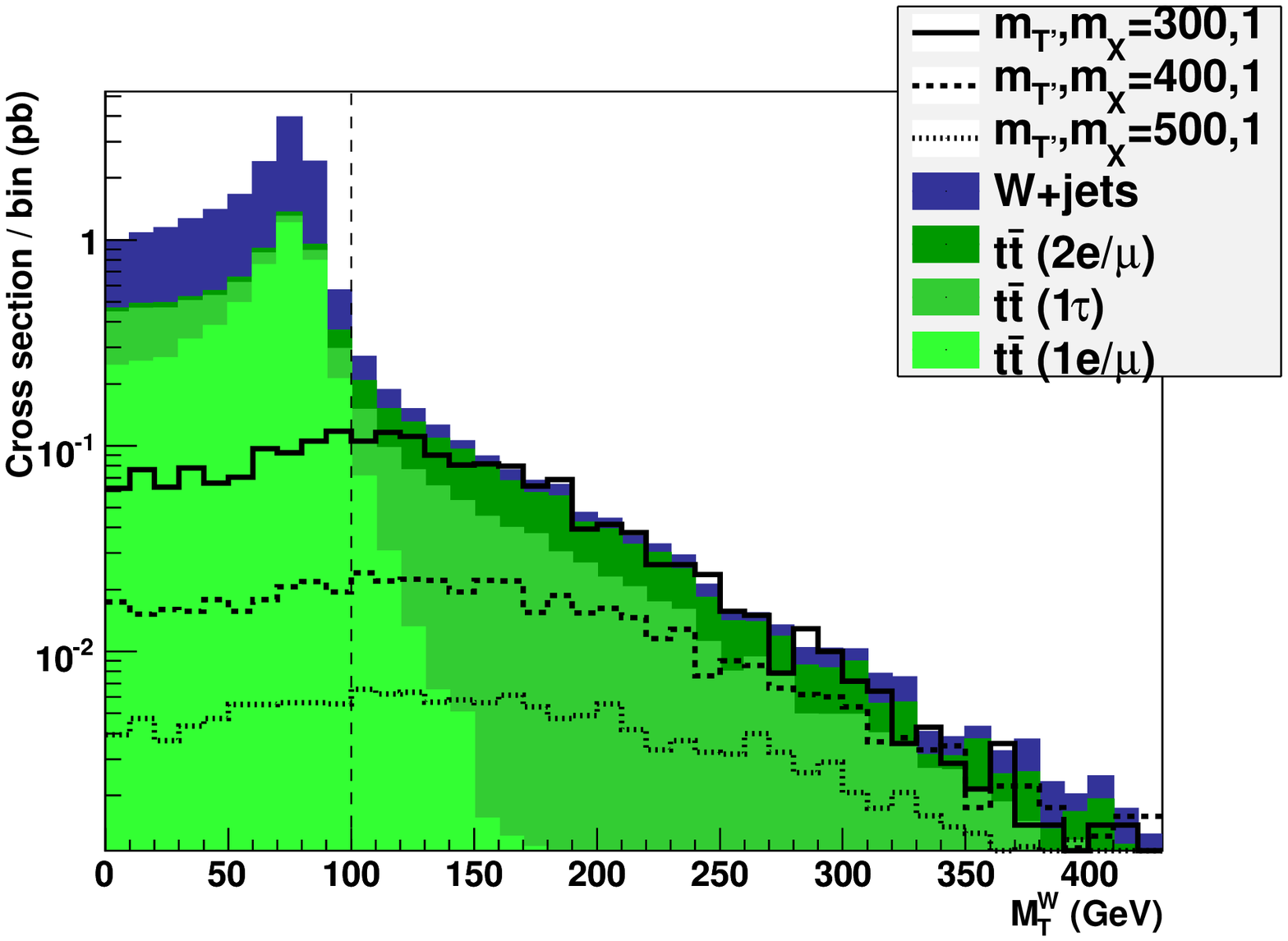}\\
\includegraphics*[width=0.48\columnwidth]{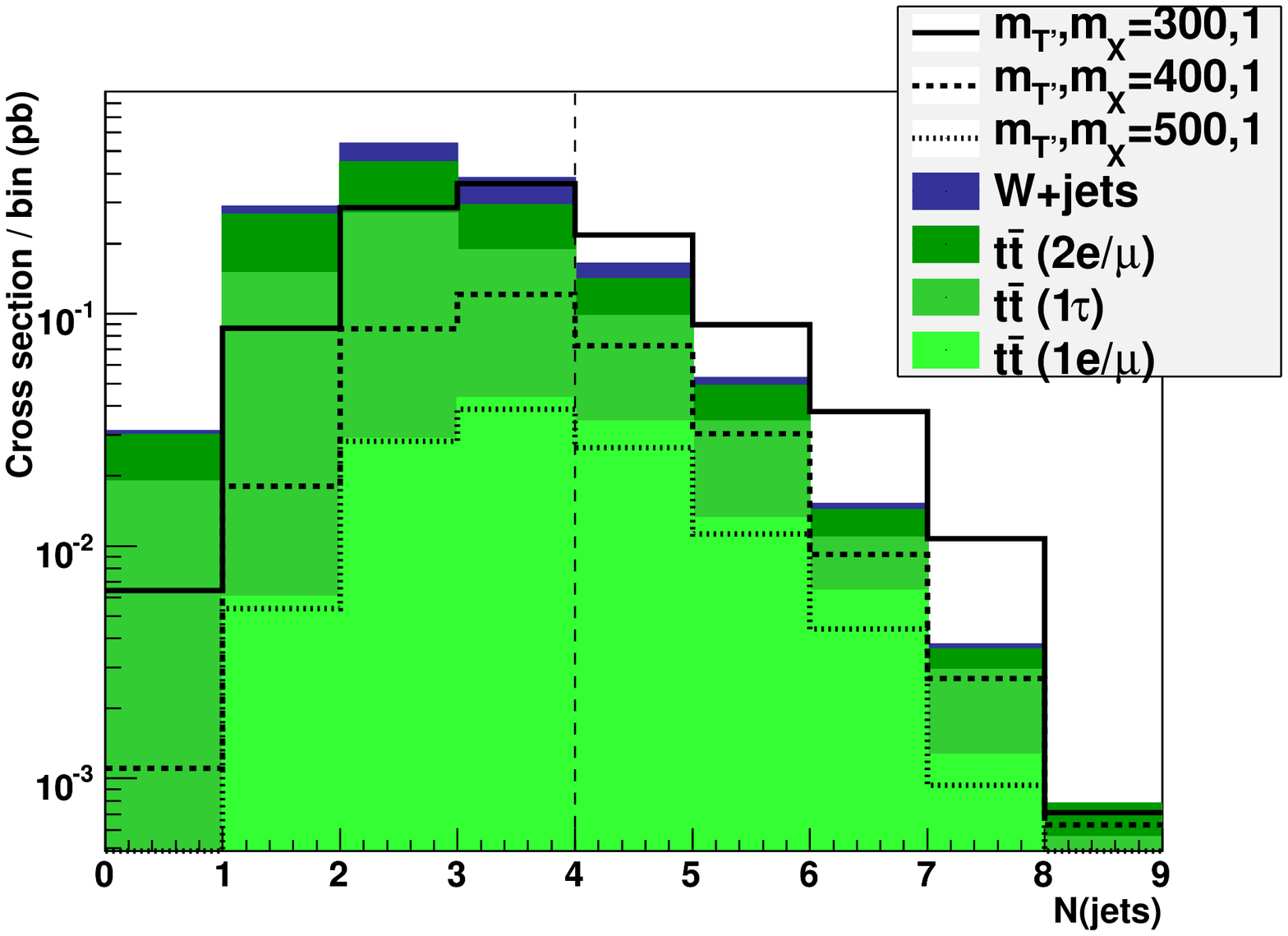}
\includegraphics*[width=0.48\columnwidth]{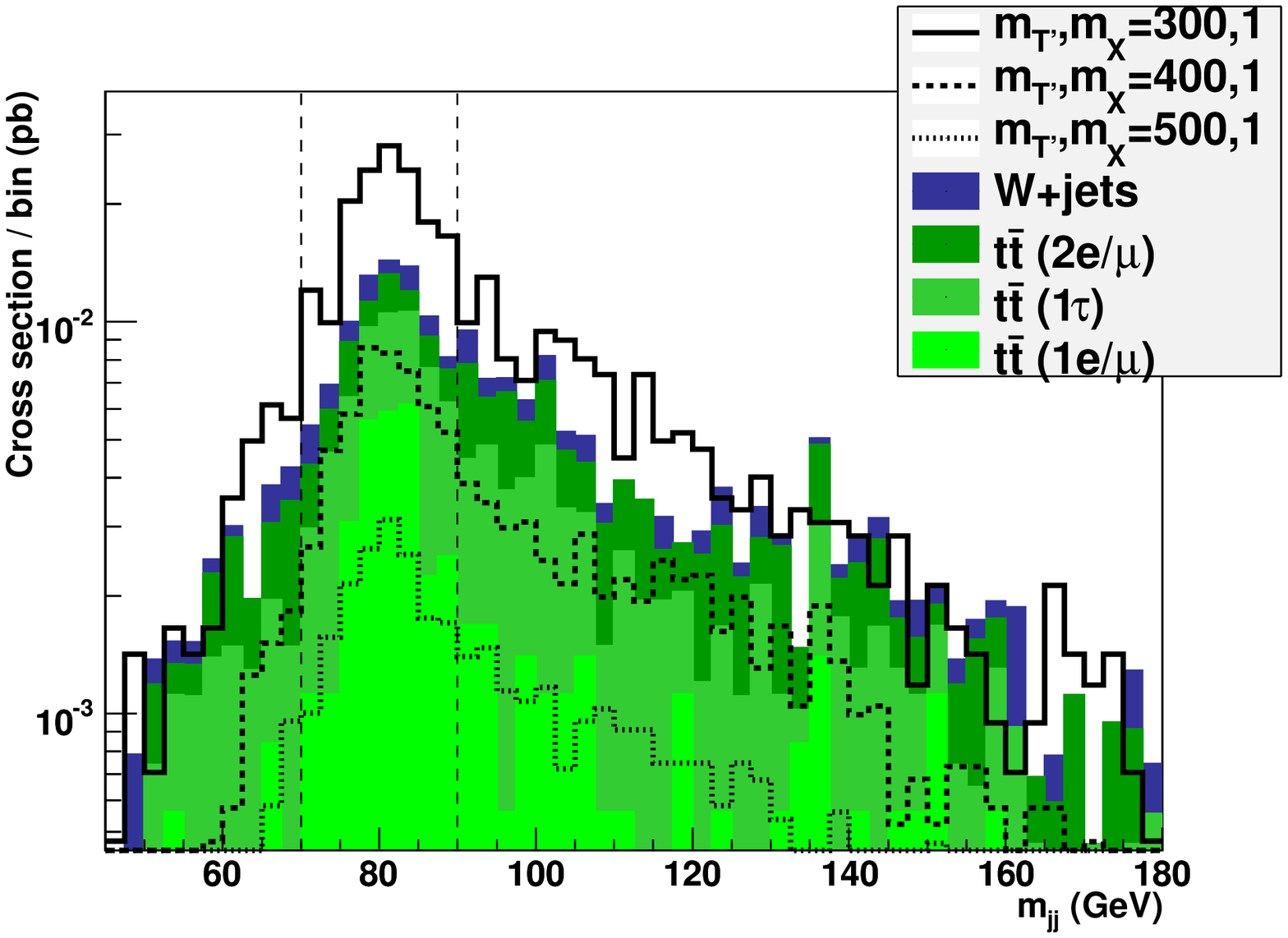}
\end{center}
\vspace*{-.25in}
\caption{\label{fig:1lep-distributions} Distributions of missing
transverse energy $\,\met$, transverse mass $m_T^W$, number
of jets $N(\text{jets})$, and jet pair invariant mass $m_{jj}$ for
signal and backgrounds for the 10 TeV LHC in the semi-leptonic
channel. Each of the observables has been plotted after the precuts
coming before it in the list, and the chosen precut has been marked by
a vertical line. For signal, the masses $(m_{T'}, m_X) = (300~\gev,
1~\gev)$, $(400~\gev, 1~\gev)$, and $(500~\gev, 1~\gev)$ have been
chosen for illustration.  The $W$ and $Z$ samples were simulated with
a cut on $\,\met>80~\gev$ and at least 3 jets in the parton-level
generation. See text for details.
}
\end{figure}

The combined background cross section after precuts is 2.4 fb for the
Tevatron, and 82 fb for the 10 TeV LHC. Typical signal efficiency
for the precuts is 2-4\% at the Tevatron and 1-2\% at the LHC.  The
cross sections after cuts, for the main backgrounds and some example
signal parameters, are found in the Appendix in
\tablesref{TeVsemi}{LHCsemi}.

\subsection{Hadronic Channel}

For the fully hadronic case, the background is mainly leptonic $W$
decays (from $W$+jets and $t\bar t\, $), where the lepton is either
missed (or non-isolated) or a $\tau$ lepton has been mistagged as a
jet, and $Z$ + jets, where the $Z$ decays to neutrinos. We therefore
expect the background to have fewer jets than the signal, which mainly
consists of fully hadronic top decays, with missing energy from the
invisible $X$ particles. To be sure to avoid QCD multi-jet background,
we also need to apply $\Delta\phi(\mpt, p_T^j)$ cuts between the
hardest jets and the missing energy. In fact, this also helps to
reduce the $W\to\tau\nu$ background, since with large $\met$ cuts, the
$W$ tends to be boosted, while the tau jet tends to be in the
direction of the missing energy. The signal is furthermore expected to
have larger $H_T=\sum |p_T^j|$ than the background.

For the fully hadronic channel, we use the following precuts:
\begin{itemize}
\setlength{\itemsep}{1pt}\setlength{\parskip}{0pt}\setlength{\parsep}{0pt}
\item No isolated electrons, muons or tau-tagged jets with $|p_T^l| >
2~\gev$.
\item Minimum missing transverse energy: $\met > 100~\gev$.
\item At least 5 jets with $|p_T^j| > 20~\gev$ (Tevatron) or $|p_T^j| >
40~\gev$ (LHC).
\item Minimum $\Delta\phi(\mpt, p_T^j)$ for the leading jets:
$\Delta\phi(\mpt, p_T^{j1}) > 90^\circ$ and $\Delta\phi(\mpt,p_T^{j2}) >
50^\circ$ (Tevatron); $\Delta\phi(\mpt, p_T^{j}) > 11.5^\circ$ for
the first, second and third leading jets (LHC).
\end{itemize}
We also use the following additional cuts to optimize the signal
significance:
\begin{itemize}
\setlength{\itemsep}{1pt}\setlength{\parskip}{0pt}\setlength{\parsep}{0pt}
\item Additional $\met$ cuts: $\met > 150, 200, 250~\gev$ (Tevatron);
$\met > 150, 200, 250, 300~\gev$ (LHC).
\item $H_T = \sum_{i=1}^5 |p_T^j|_i$ cuts: $H_T > 300, 350, 400~\gev$
(Tevatron); $H_T > 400, 500~\gev$ (LHC).
\item At least 6 jets with $|p_T^j| > 20~\gev$ (Tevatron) or $|p_T^j| >
40~\gev$ (LHC).
\end{itemize}

As discussed above, the relevant backgrounds for the fully hadronic
channel are $t\bar t$, leptonically-decaying $W^\pm$ + jets, and
$Z\to\nu\bar\nu$ + jets. For completeness, we also simulated $t\bar t
Z$, but this is negligible because of its small cross section. Among
the $t\bar t$ decay modes, the dominant background is from decays with
at least one tau lepton, followed by the semi-leptonic decay to
electron or muon (where the lepton is either missed or non-isolated).

\begin{figure}[tb]
\begin{center}
\includegraphics*[width=0.48\columnwidth]{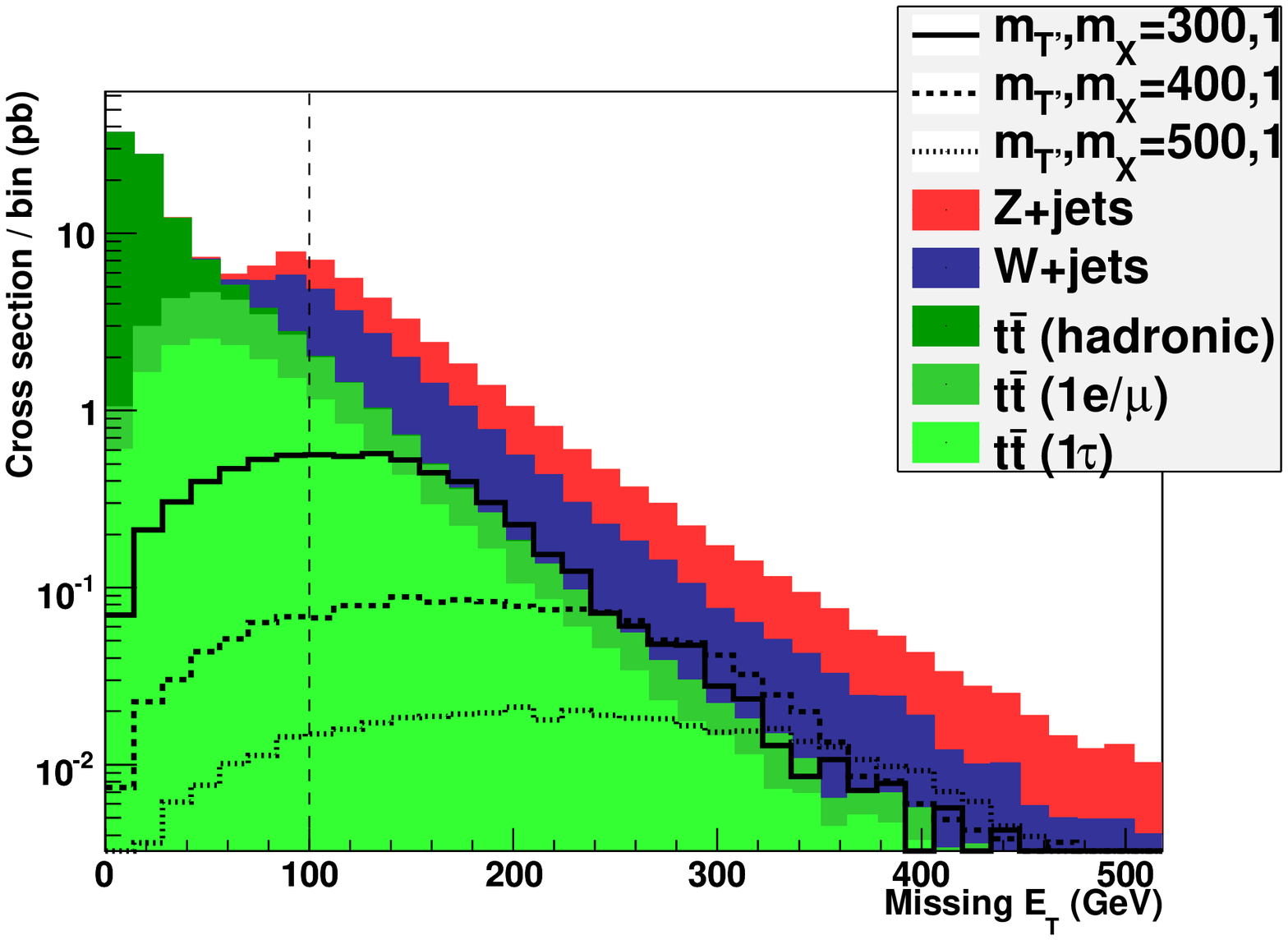}
\includegraphics*[width=0.48\columnwidth]{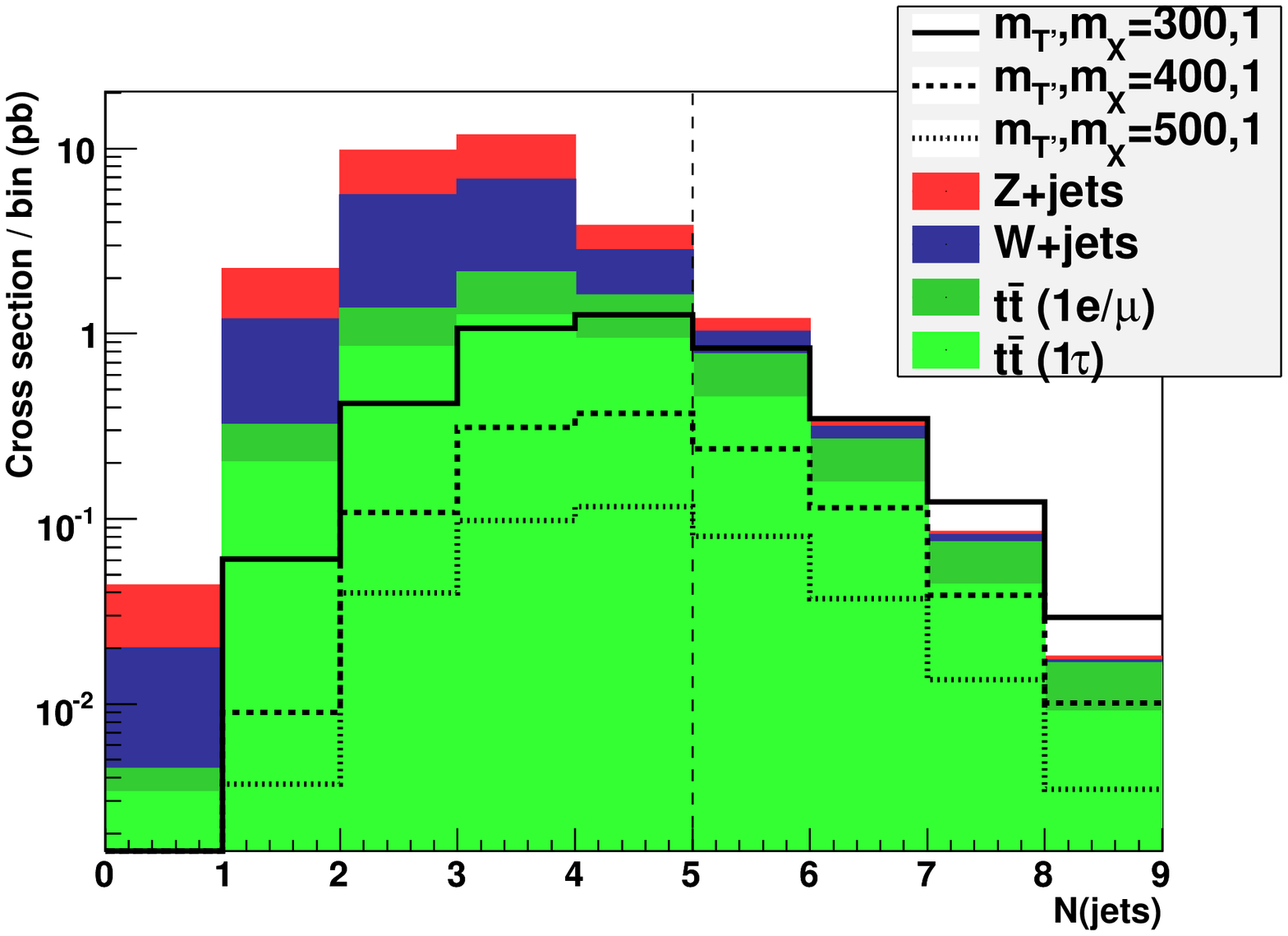}\\
\includegraphics*[width=0.48\columnwidth]{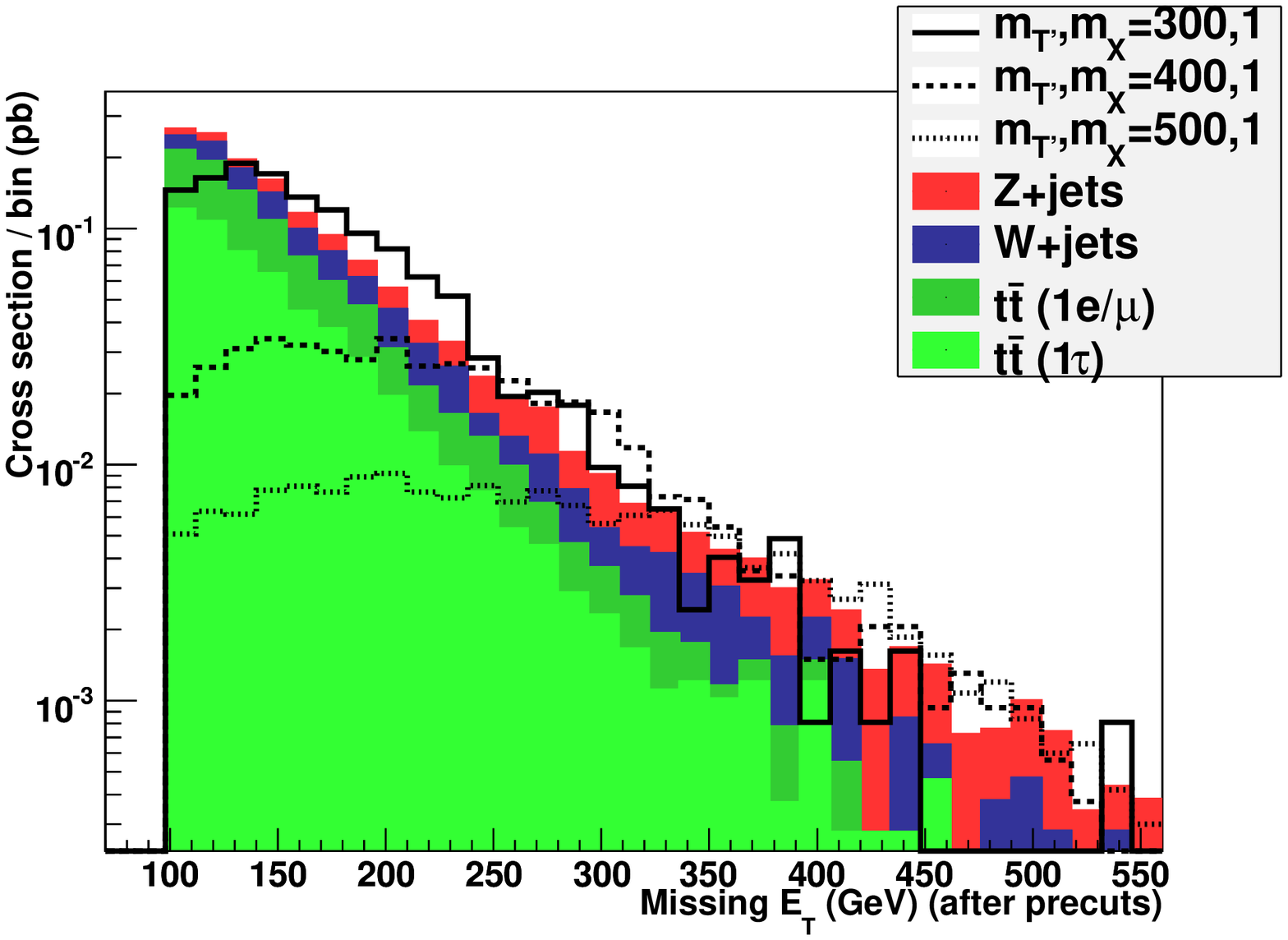}
\includegraphics*[width=0.48\columnwidth]{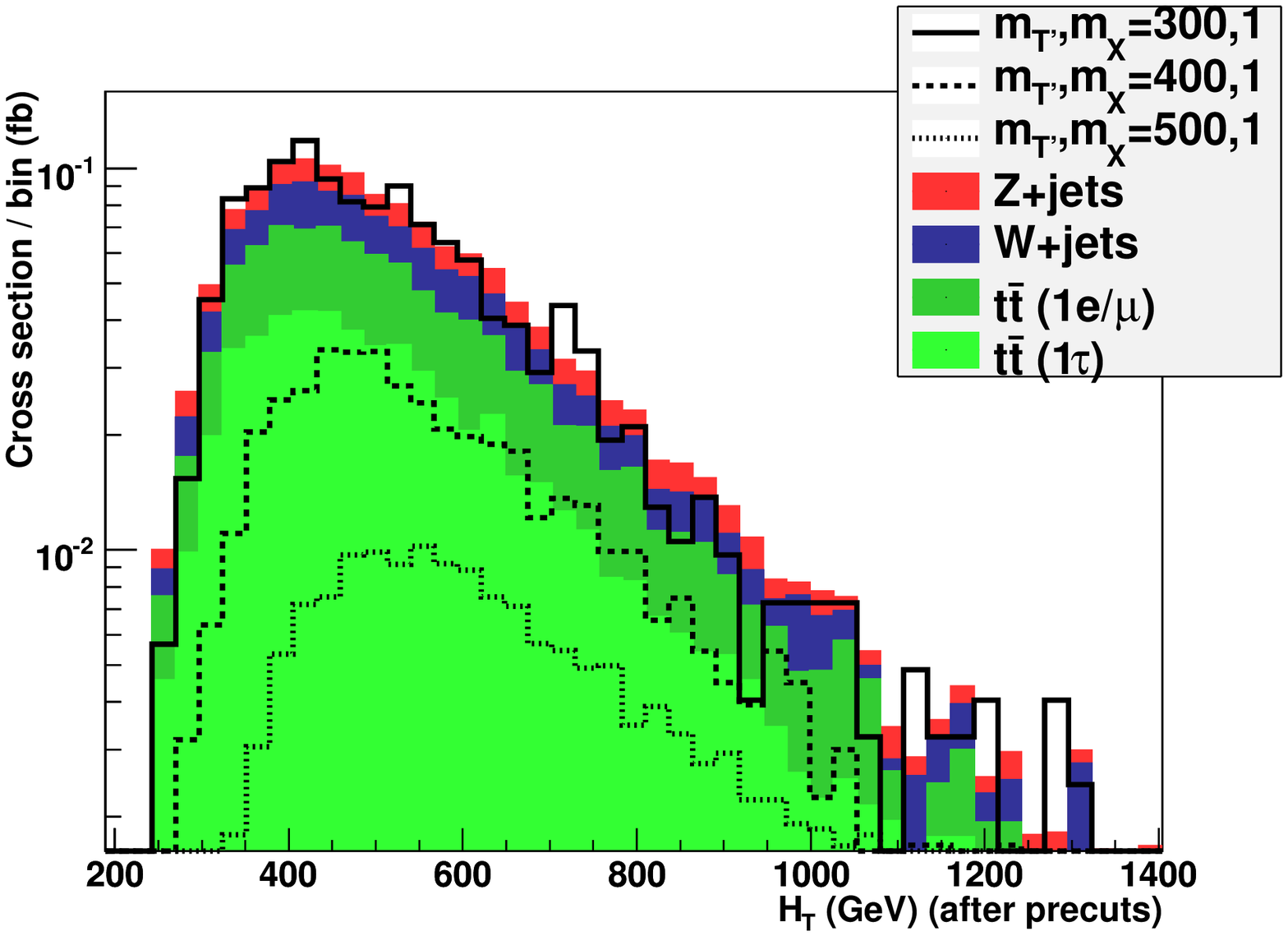}
\end{center}
\vspace*{-.25in}
\caption{\label{fig:nolep-distributions} $\met$, $N(\text{jets})$, and
$H_T$ distributions for signal and backgrounds for the 10 TeV LHC in
the hadronic channel. The top two panels show distributions of
$\,\met$ and $N(\text{jets})$ after the previous cuts in the precut
table, with the position of the precut marked with a vertical dashed
line.  The lower two panels show distributions of $\,\met$ and $H_T$
after all precuts.  The hadronic top contribution is negligible after
the $\,\met>100~\gev$ cut and has therefore been omitted in the
remaining plots. For signal, the masses $(m_{T'}, m_X) = (300~\gev,
1~\gev)$, $(400~\gev, 1~\gev)$, and $(500~\gev, 1~\gev)$ have been
chosen for illustration. The $W$ and $Z$ samples were simulated with a
cut on $\,\met > 80~\gev$ and at least 3 jets in the parton-level
generation.  See text for details.  }
\end{figure}

Distributions for $\met$, $N(\text{jets})$, and $H_T$ for both signal
and backgrounds in the hadronic channel at the 10 TeV LHC are shown in
\figref{nolep-distributions}.  The top two panels show $\met$ and
$N(\text{jets})$ plotted after the cuts coming before it in the list,
and the position of the precut is marked with a vertical dashed
line. The bottom two panels are $\met$ and $H_T$ distributions plotted
after precuts. For clarity, we have split the $t\bar t$ background
into components: fully hadronic decay (negligible after $\met$ cut),
decays with at least one tau lepton, semi-leptonic decays (to electron
or muon), and purely leptonic decays (which are negligible with these
cuts).  The corresponding distributions for the Tevatron are
qualitatively similar.

After precuts for the hadronic channel, the combined background cross
section is 21 fb for the Tevatron and 1.4 pb for the 10 TeV LHC. The
signal efficiency of the precuts is 9-20\% at the Tevatron and 8-13\%
at the LHC. A table of cross sections after cuts for backgrounds and
some signal points may be found in the Appendix in
\tablesref{TeVhad}{LHChad}.

The main remaining backgrounds after precuts for both the
semi-leptonic and hadronic channels include tau leptons. One reason
for this is that a tau lepton is often mistagged as a jet, which
therefore adds significantly to the fully hadronic background with
large $\met$ (in particular for the hadronic channel). It would be
interesting to see an experimental study of whether an anti-tau tag
could be effective in further suppressing these backgrounds, while
keeping a good signal efficiency.  This might be of significant
importance for any new physics with signatures consisting of jets and
missing energy.

\section{Discovery and Exclusion Reach from Tevatron and Early LHC Data}
\label{sec:discovery}

We now determine the discovery and exclusion reach for $T'$ at the
Tevatron and the 10 TeV LHC. For each parameter point $(m_{T'},m_X)$,
we use the optimum cut (after precuts) that gives the best signal
significance, with the additional requirements that $S/B > 0.1$ and
more than two signal events are observed. Given the small number of
signal and background events after cuts, we have used Poisson
statistics, rather than assuming Gaussian distributions, for both
signal and backgrounds.

\Figref{exclusion-TeV} shows the 95\% CL Tevatron exclusion contours
for both the semi-leptonic and hadronic channels and integrated
luminosities of 2, 5, 10, and $20~\ifb$.  Even with just $2~\ifb$,
exclusion limits of $m_{T'} > 340~\gev$ (semi-leptonic mode) and
$m_{T'} > 380~\gev$ (hadronic mode) can be reached, which already
extend into the interesting mass range consistent with current direct
search bounds and precision electroweak data.  With a combined
integrated luminosity of $20~\ifb$ at the end of Tevatron running, a
reach of up to 455 GeV for the hadronic channel can be achieved.

\begin{figure}[tb]
\begin{center}
\includegraphics*[width=0.48\columnwidth]{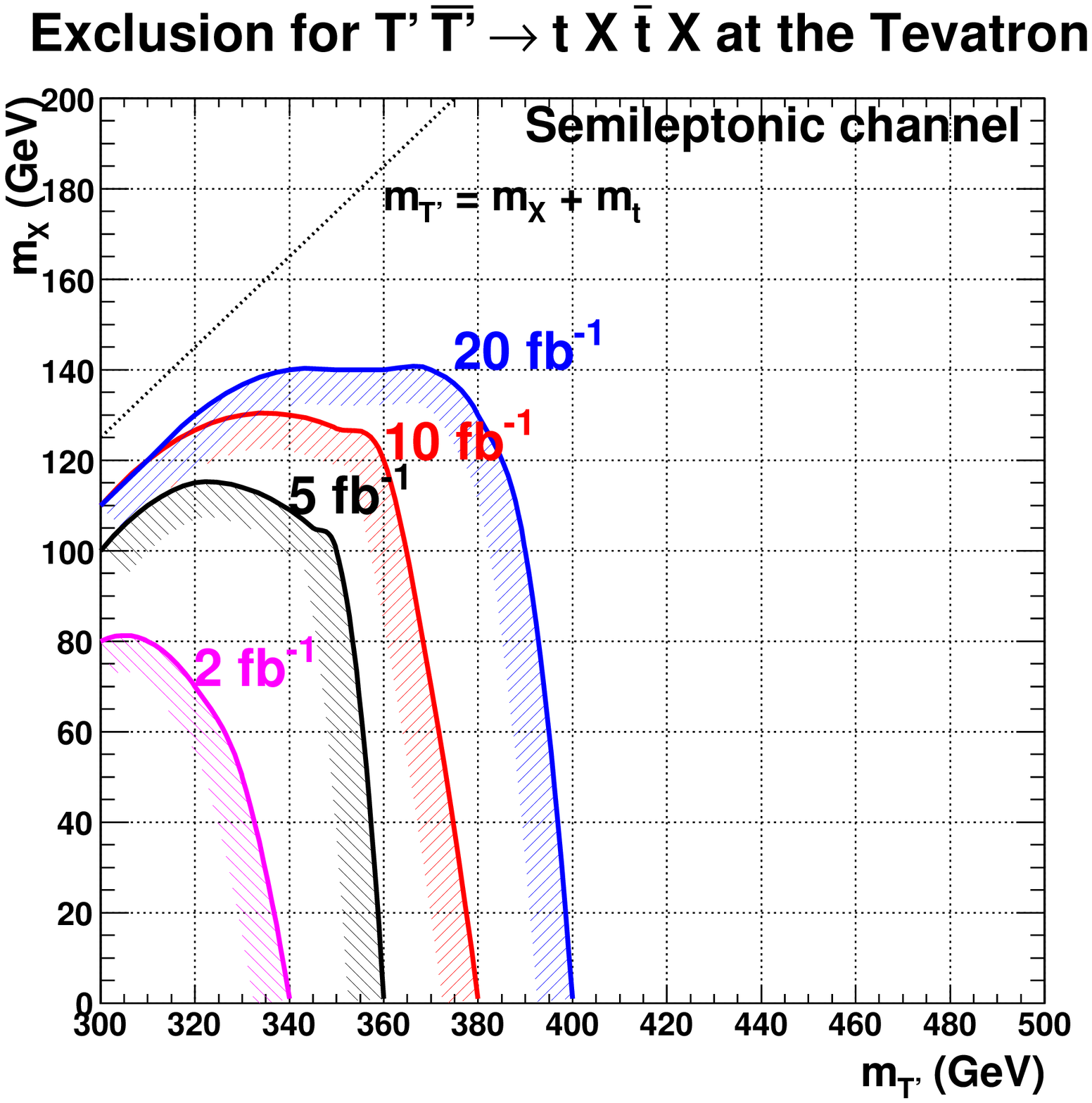}
\includegraphics*[width=0.48\columnwidth]{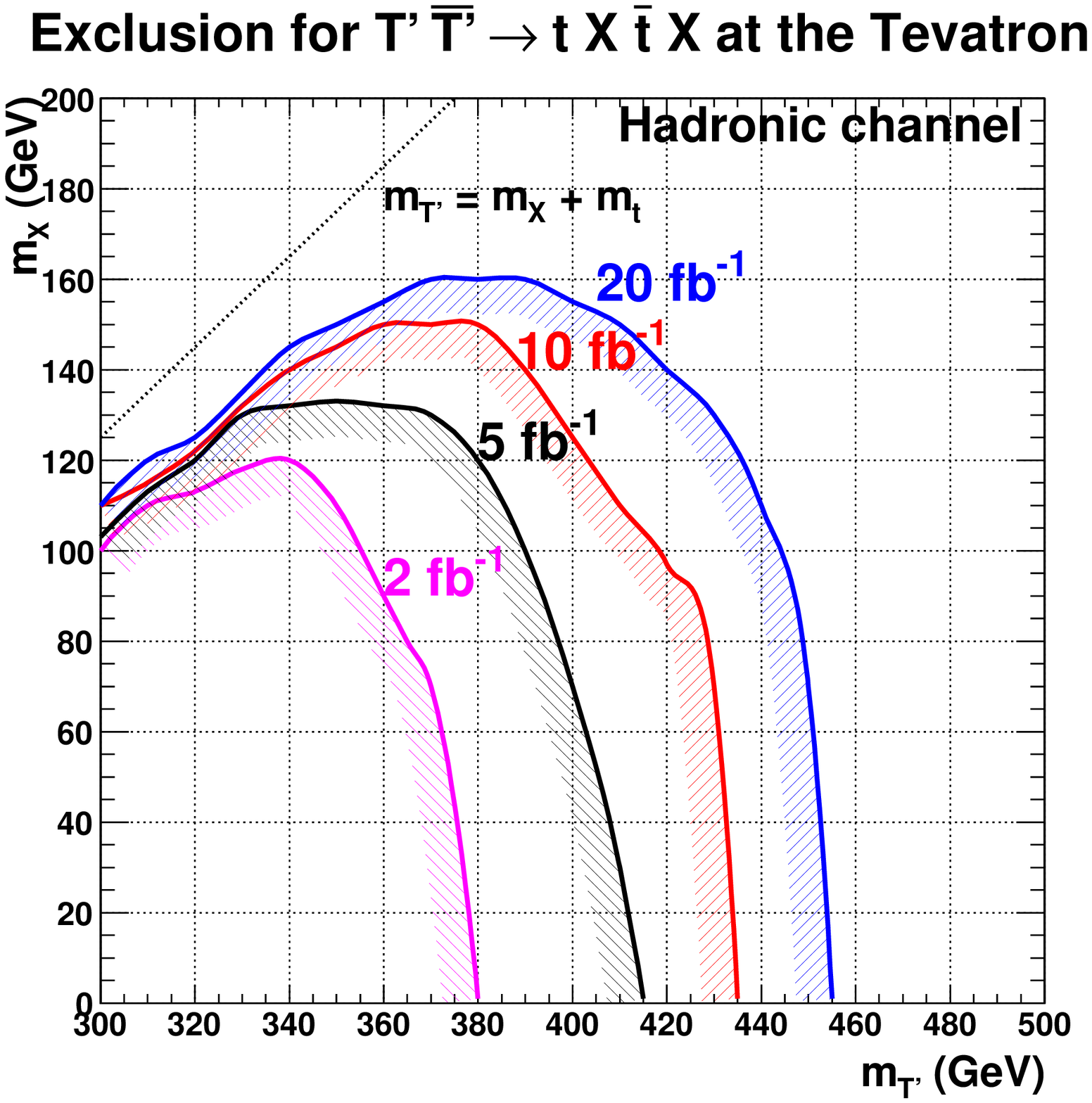}
\end{center}
\vspace*{-.25in}
\caption{\label{fig:exclusion-TeV} 95\% CL Tevatron exclusion contours
for the semi-leptonic channel (left) and the hadronic channel (right)
for integrated luminosities 2, 5, 10, and $20~\ifb$. For each point in
parameter space, the cut with the best significance has been chosen.}
\end{figure}

The reach in $m_{T'}$ is almost independent of $m_X$ for small to
medium $m_X$.  However, when $m_X$ approaches the on-shell decay
threshold of $m_{T'}-m_t$, the reach is limited since the top and $X$
are produced nearly at rest in the $T'$ rest frame, and the
$T'\bar{T'}$ system therefore needs a transverse boost for the $X$
particles to produce large missing transverse momentum.  This leads to
the dip in the exclusion curves at $m_{X}$ close to $m_{T'}-m_t$, and
indeed there is no exclusion reach at the Tevatron for $m_{T'}-m_t-m_X
\alt 15~\gev$.  For $20~\ifb$ integrated luminosity and $m_{T'}$
between 370 and 390 GeV, $m_X$ could be excluded up to 160 GeV at 95\%
CL using the hadronic mode.  For smaller $m_{T'}$, the reach in $m_X$
is decreased due to the softness of the $X$ particle distributions,
while for larger $m_{T'}$, it is decreased because of the small $T'
\bar{T}'$ production cross section.

\Figref{discovery-TeV} shows the 3$\sigma$ (Gaussian
equivalent\footnote{By Gaussian equivalent, we mean that we have
converted the one-sided Poisson probability into the equivalent
$\sigma$ deviation in a two-sided Gaussian distribution, which is more
commonly used in the literature.}) Tevatron discovery contours for
both the semi-leptonic and hadronic channels for integrated
luminosities of 2, 5, 10, and $20~\ifb$.  A 3$\sigma$ signal could be
observed for $m_{T'}<360~\gev$ and $m_X\alt 110~\gev$ in the
semi-leptonic channel with $20~\ifb$ integrated luminosity.  The
hadronic channel is more promising.  With $5~\ifb$ integrated
luminosity, a reach in $m_{T'}$ up to 360 GeV could be achieved when
$m_X$ is not too large.  With $20~\ifb$ integrated luminosity, the
reach is extended to 400 GeV for $m_X$ up to about 80 GeV.  For larger
$m_X$, the reach in $m_{T'}$ decreases.

\begin{figure}[tb]
\begin{center}
\includegraphics*[width=0.48\columnwidth]{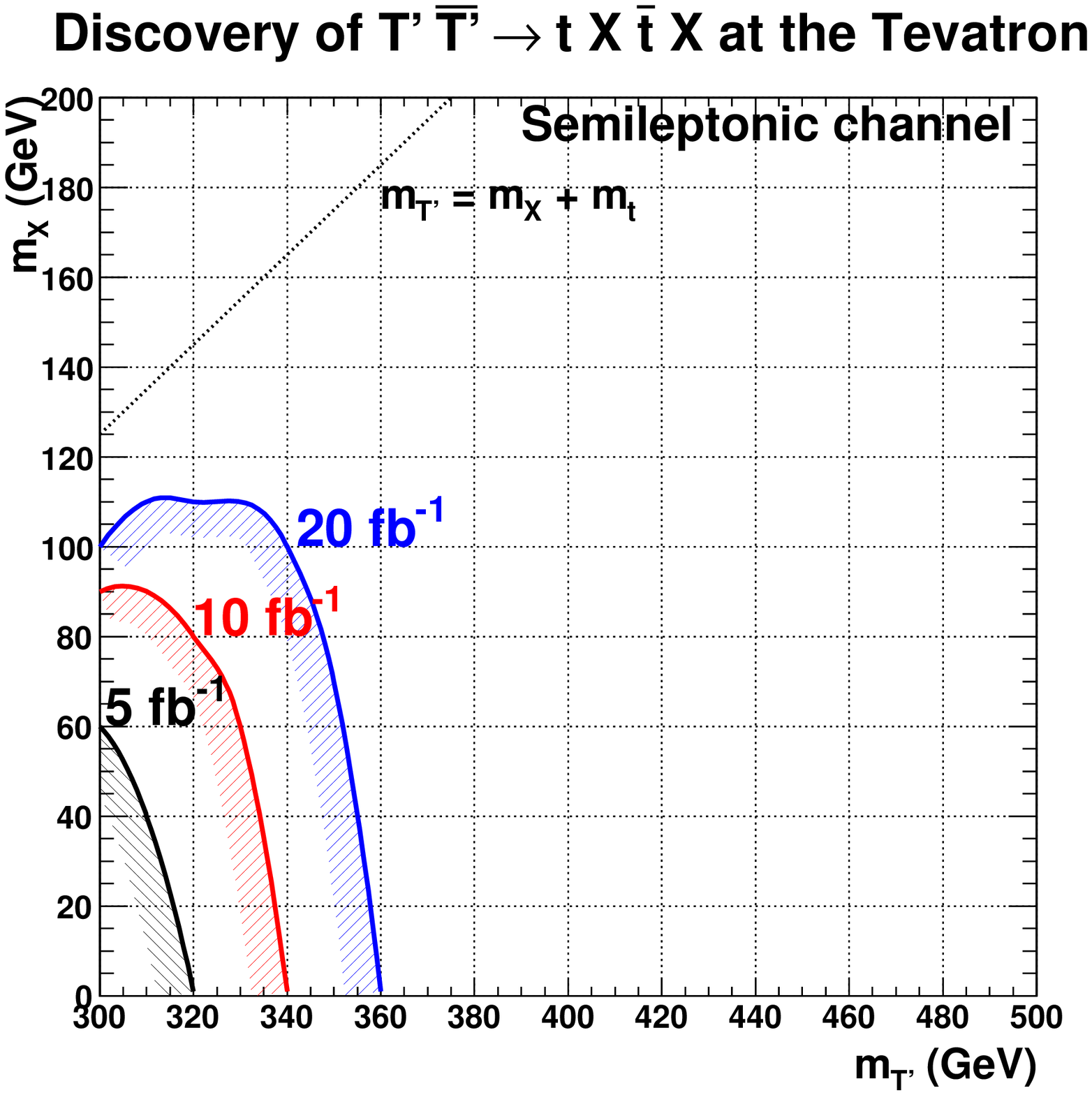}
\includegraphics*[width=0.48\columnwidth]{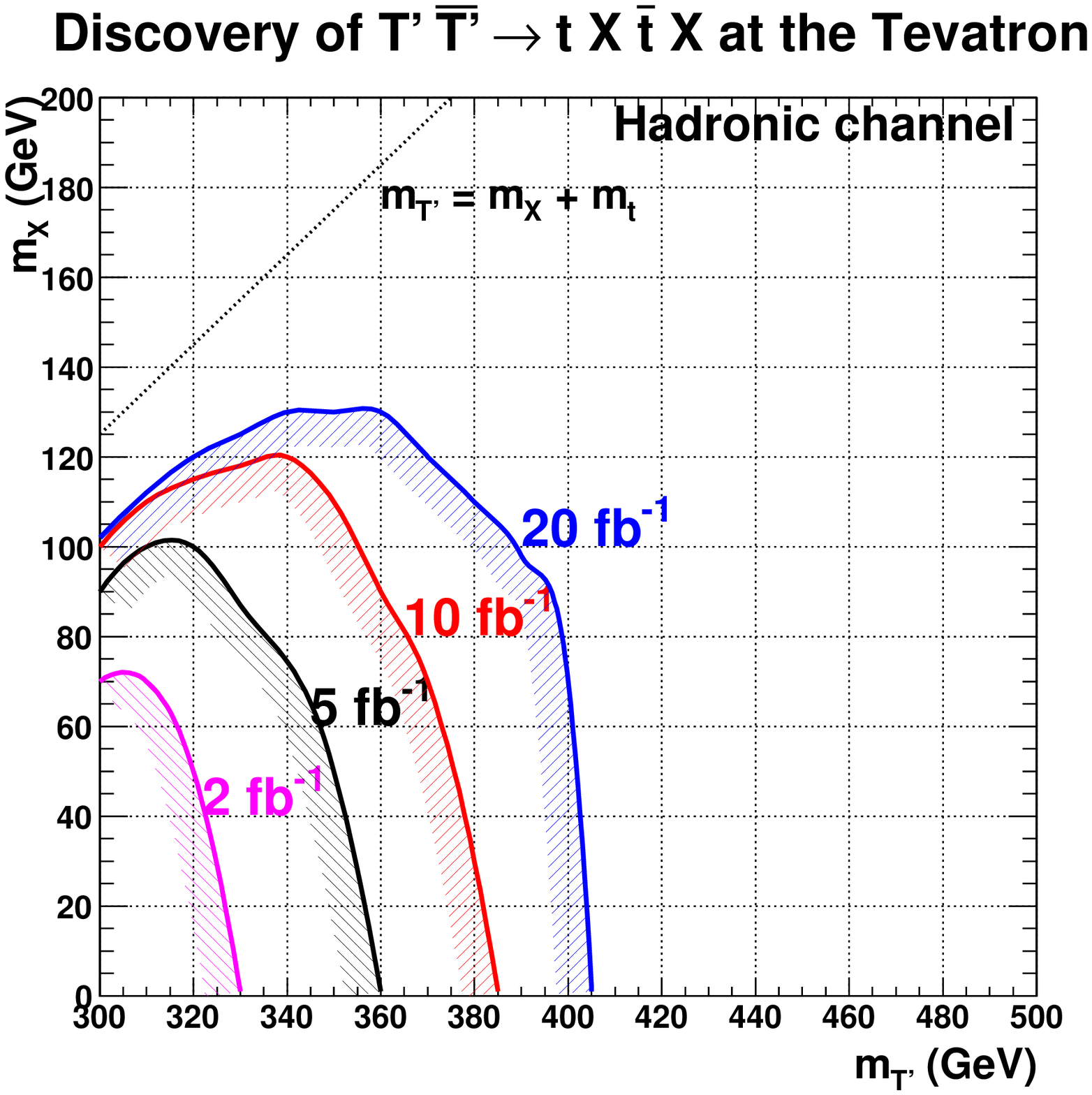}
\end{center}
\vspace*{-.25in}
\caption{\label{fig:discovery-TeV} 3$\sigma$ (Gaussian equivalent)
Tevatron discovery contours for the semi-leptonic channel (left) and
the hadronic channel (right) for integrated luminosities 2, 5, 10,
and $20~\ifb$. For each point in parameter space, the cut with the
best significance has been chosen.}
\end{figure}

\Figref{exclusion-LHC10} shows the 95\% CL exclusion contours for a 10
TeV early LHC run, in the semi-leptonic and hadronic channels for
integrated luminosities 100, 200, and $300~\ipb$.  With just
$100~\ipb$, the LHC exclusion reach for $m_{T'}$ exceeds the Tevatron
exclusion reach with $20~\ifb$ luminosity.  Exclusions of $m_{T'}$ up
to 490, 520, and 535 GeV could be achieved with 100, 200, and
$300~\ipb$ integrated luminosity for the semi-leptonic channel.  The
exclusion region for the hadronic channel covers almost the entire
interesting mass parameter space with $300~\ipb$ luminosity.  Note
that at the LHC, we could tolerate much smaller $m_{T'}-m_X$; in
particular, we start probing the off-shell decay region $T' \to
t^*X\to bWX$ for $m_{T'}-m_X < m_t$.

\begin{figure}[tb]
\begin{center}
\includegraphics*[width=0.48\columnwidth]{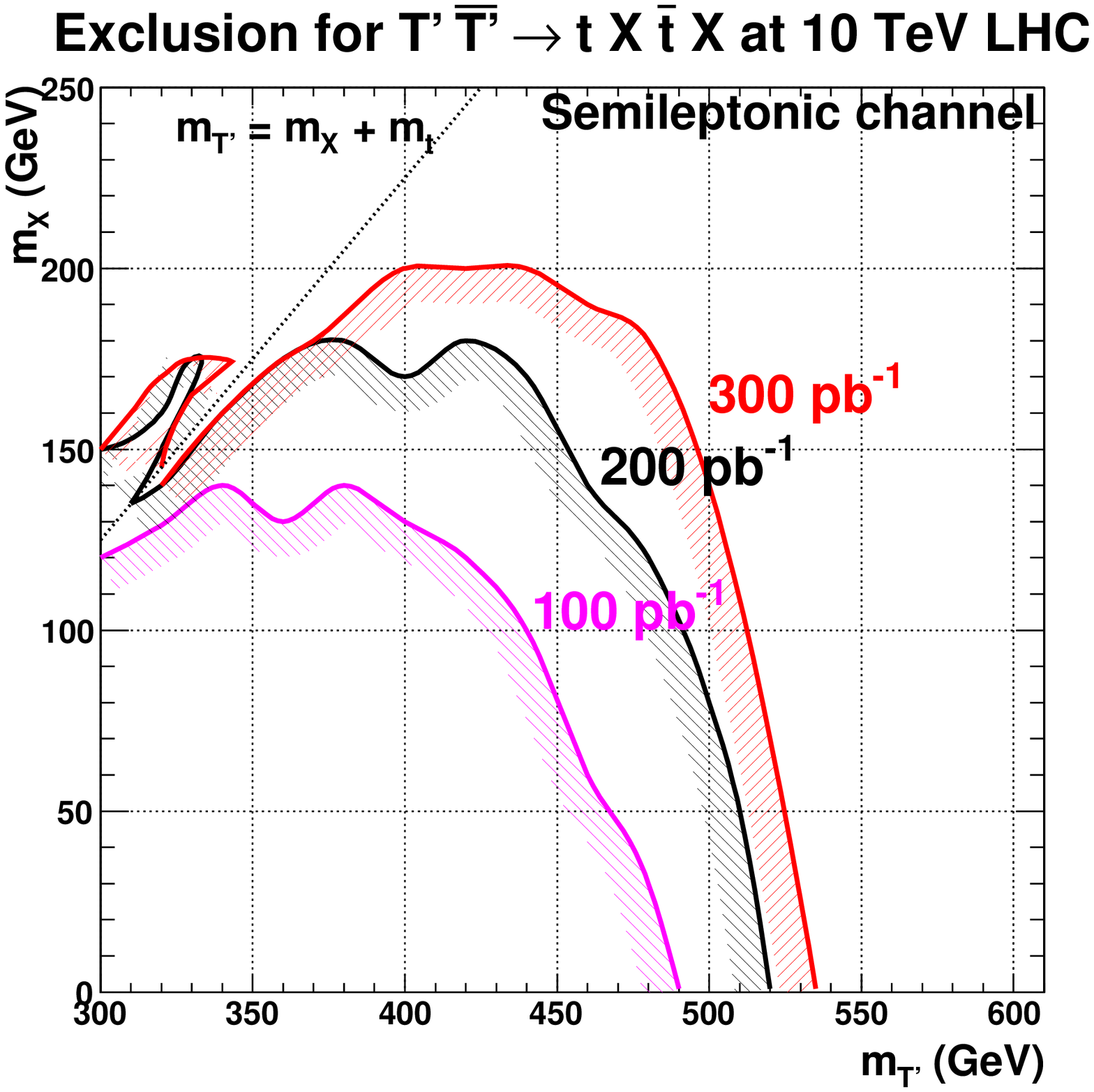}
\includegraphics*[width=0.48\columnwidth]{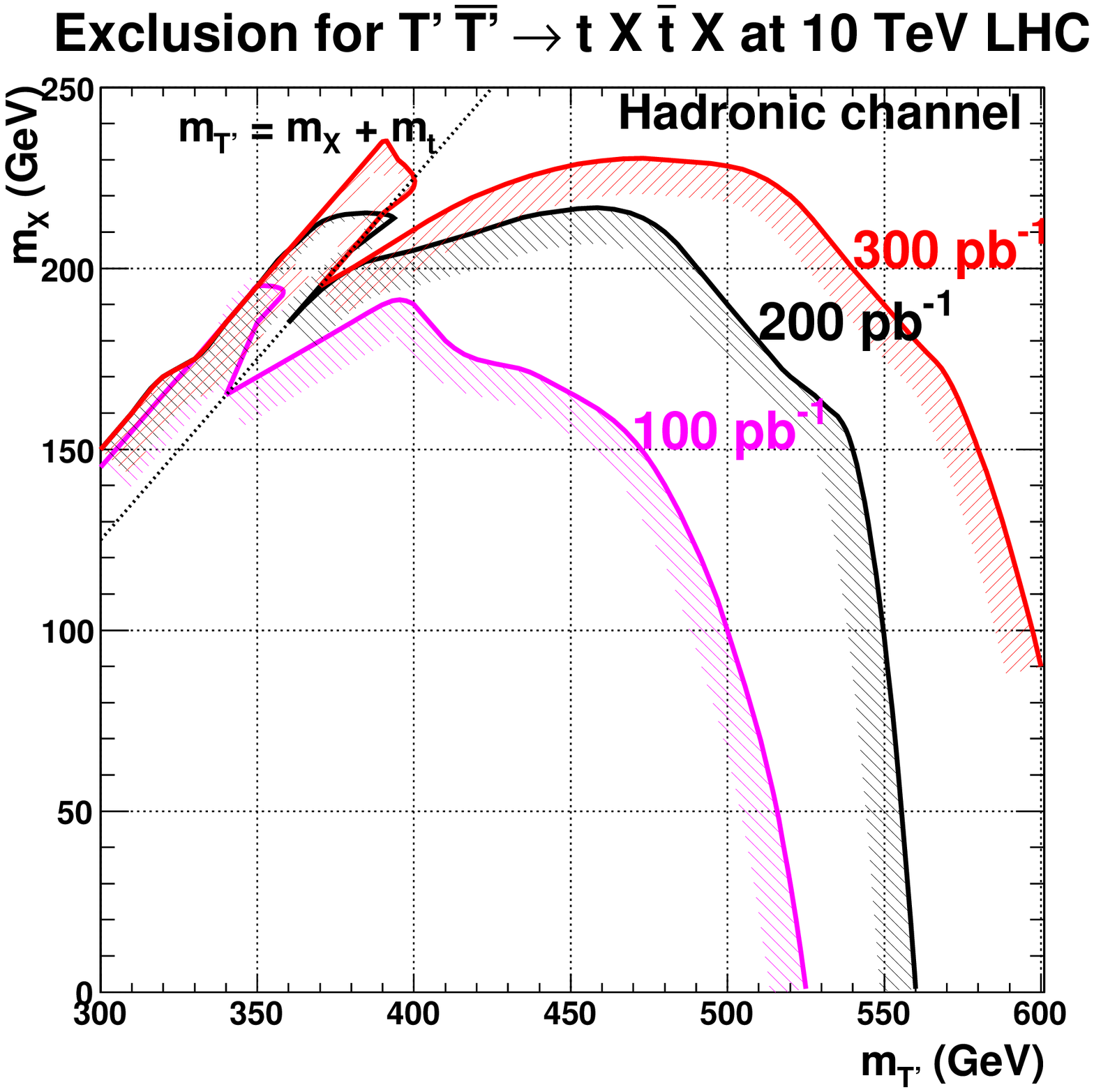}
\end{center}
\vspace*{-.25in}
\caption{\label{fig:exclusion-LHC10} 95\% CL exclusion contours for a
10 TeV LHC run in the semi-leptonic channel (left) and the hadronic
mode (right), for integrated luminosities 100, 200, and
$300~\ipb$. For each point in parameter space, the cut with the best
significance has been chosen.}
\end{figure}

\Figref{discovery-LHC10} shows the 3$\sigma$ (Gaussian equivalent)
discovery contours for a 10 TeV LHC run, in the semi-leptonic and
hadronic channels for integrated luminosities 100, 200, and
$300~\ipb$.  Although the reach in both $m_{T'}$ and $m_X$ is limited
for the semi-leptonic mode, the hadronic channel could provide a
3$\sigma$ signal for $m_{T'} \alt 490~\gev$ and $m_X \alt 170~\gev$
with $300~\ipb$ luminosity.  We might also observe a positive signal
for $m_X$ up to about 170 GeV in the off-shell decay region ($m_{T'} -
m_X < m_t$) for $m_{T'} \alt 330~\gev$.

\begin{figure}[tb]
\begin{center}
\includegraphics*[width=0.48\columnwidth]{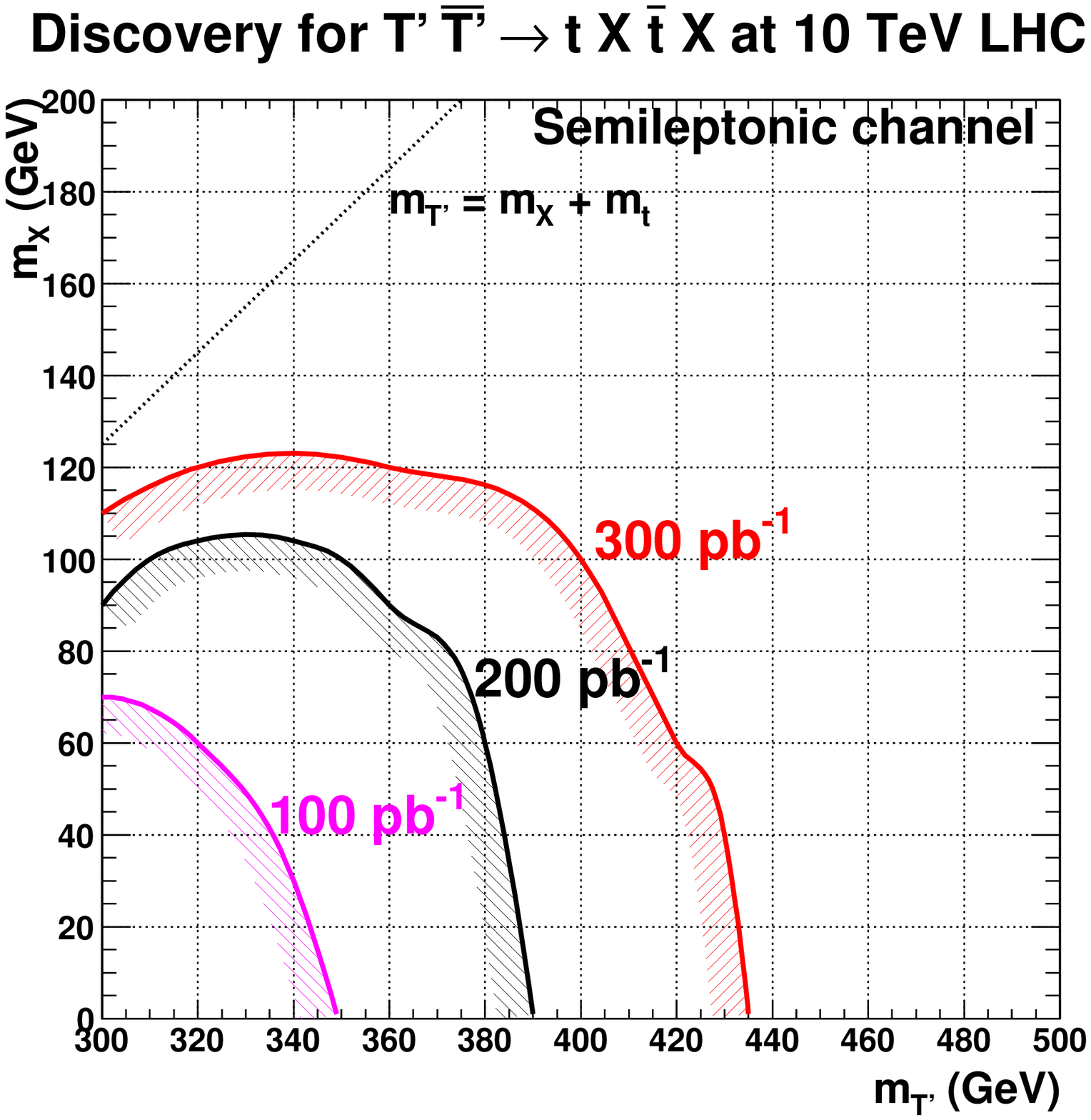}
\includegraphics*[width=0.48\columnwidth]{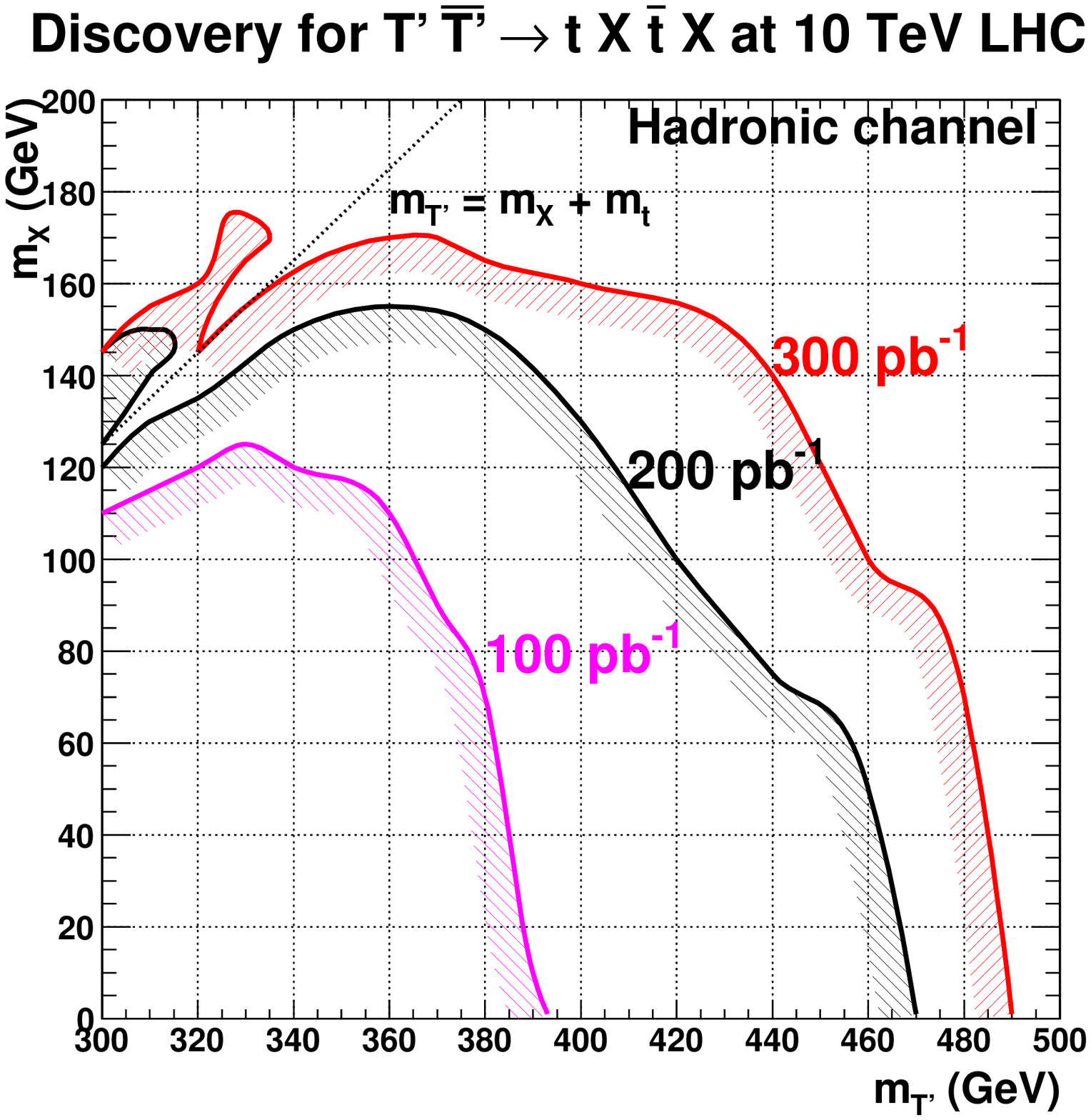}
\end{center}
\vspace*{-.25in}
\caption{\label{fig:discovery-LHC10} 3$\sigma$ (Gaussian equivalent)
discovery contours for a 10 TeV LHC run, in the semi-leptonic channel
(left) and the hadronic channel (right), for integrated luminosities
100, 200, and $300~\ipb$. For each point in parameter space, the cut
with the best significance has been chosen. }
\end{figure}

It is clear from the discovery and exclusion contours, both for the
Tevatron and the LHC, that the fully hadronic channel has considerably
larger reach than the semi-leptonic channel, for reasons enumerated in
\secref{simulation}. In this channel, the full, currently viable,
region in parameter space can be excluded at a 10 TeV LHC
run.\footnote{The results obtained here can be readily translated to
an LHC run at 7 TeV, by multiplying the integrated luminosities needed
by roughly a factor of 3.  This approximation accounts for the
difference in cross sections at different center of mass energies,
assuming that the cut efficiencies for both the signal and backgrounds
do not change significantly.} In case both channels are visible, they
can be used to distinguish between different model and mass
hypotheses.

\section{Conclusions}
\label{sec:conclusions}

We have considered the prospects for hadron colliders to pair produce
exotic 4th generation quarks that decay directly to a pair of dark
matter particles and SM particles.  Although we have a particular
interest in the WIMPless dark matter scenario~\cite{Feng:2008ya}
(including a specific example~\cite{Feng:2008dz} that can potentially
explain the DAMA annual modulation result), this scenario is motivated
on quite general grounds, and, with minor modifications, our analysis
applies to many other dark matter scenarios and other new physics
models.

We have focused on the up-type 4th generation quark $T'$.  $T'$ pair
production leads to $T' \bar{T}' \to t \bar{t} X X$, and we have then
analyzed the semi-leptonic and fully hadronic channels.  The fully
hadronic channel (vetoing events with leptons) seems to be the most
efficient, because of the large branching fraction and the reduction
in SM background with large $\met$.

Existing constraints require $300~\gev \alt m_{T'} \alt 600~\gev$,
where the lower bound comes from direct searches, and the upper bound
is from perturbativity.  We have found that there are bright prospects
for probing exotic 4th generation quarks in this mass window at the
Tevatron and in early data from the LHC.  For models with $m_X \alt
120~\gev$, the discovery of new physics is possible at the Tevatron
with $\sim 10~\ifb$ of luminosity, while for $m_X \alt 170~\gev$ the
discovery of new physics may be possible at the LHC with $\sim
300~\ipb$.  In particular, with $\sim 300~\ipb$ of data, the LHC
should be able to discover almost all of the relevant parameter space
with $m_X \alt 10~\gev$, where WIMPless models can explain the DAMA
and CoGeNT results.  Conversely, if no signal is seen in $300~\ipb$,
the entire mass range consistent with current bounds and
perturbativity will be excluded.

Of course, although an exclusion definitively excludes the model, a
discovery will only be a discovery of a multi-jet (+ lepton) + $\met$
signal. Considerably larger integrated luminosity would be needed to
identify the signal as $t\bar t + \met$, and it is an even harder
problem to determine if the new physics really is $T' \bar{T'}$
production, with decay to top quarks and dark matter.  Such a
discovery analysis would require a good identification of the decaying
top quarks as well as spin and mass determinations of the $T'$ and $X$
particles, and would require significant amounts of data from the LHC
(and be beyond the capabilities of the Tevatron).  Exotic 4th
generation quark decays are not the only processes that give a signal
of top quark pairs plus missing transverse energy.  In particular,
this is a typical signature for supersymmetry, for example, from stop
pair production followed by the decay $\tilde t \to t \tilde
\chi_1^0$.  However, it should be possible to distinguish these
possibilities with more LHC data.  Since the 4th generation quark is a
fermion, it has a higher production cross-section than squarks with a
similar mass.  Moreover, squarks could also have more complicated
decay chains that would be absent in $T'$ decay.  There have already
been studies on how to distinguish supersymmetric signals from other
new physics with similar signals~\cite{Datta:2005zs,MeadeDW,HanGY},
and it would be worthwhile to perform a more detailed analysis of how
one would distinguish exotic 4th generation quarks from squarks.  It
is also worth noting that the process $pp \to T' \bar{T'} \to X \bar
X$ + jets would be well suited for analysis using the $m_{T2}$
kinematic variable~\cite{mt2}.

This analysis has focused on pair production of the up-type 4th
generation quarks $T'$.  Precision electroweak constraints imply that
the down-type 4th generation quark $B'$ must be fairly degenerate with
the $T'$, and so $B'\bar B' \to b \bar b X \bar X $ should also be
accessible.  This will likely be a more difficult signal to extract
from early data, since it requires a good understanding of
$b$-tagging, and we would expect large QCD backgrounds.  But in the
event of a discovery, such an analysis will be immensely useful in
understanding the underlying physics.

There is also an interesting complementarity between the collider
studies of dark matter considered here, and direct or indirect
detection strategies.  For example, one may consider WIMPless dark
matter models in the limit of small $\lambda$.  In this limit the
cross-sections for dark matter-nucleon scattering and for dark matter
annihilation are small, and direct or indirect dark matter searches
will be unsuccessful.  But the production cross section for $T'
\bar{T'}$ pairs is controlled by QCD, independent of the Yukawa
coupling $\lambda$.  So for models in this limit of parameter space,
hadron colliders may provide the only direct evidence for the nature
of dark matter.  Interestingly, at small $\lambda$, the 4th generation
quarks are long-lived.  This may result in displaced decay vertices,
and a sufficiently long-lived 4th generation quark may even hadronize
and reach the detector.  There has already been significant study of
detection strategies for long-lived exotic
hadrons~\cite{Drees:1990yw}, and these results should be directly
applicable to the case when the $T'$ travels a macroscopic distance in
the detector. It would be interesting to investigate further how to
determine the nature of such semi-stable color triplet particles.

Our results for a 10 TeV LHC run can be approximately translated to
the alternative of an extended 7 TeV run. In this case, a coverage
corresponding to the $300~\ipb$ quoted in this study should be
attainable for less than about $1~\ifb$.

Finally, and of particular interest, the annual modulation signal of
DAMA~\cite{DAMAlowmass} has been recently supported by unexplained
events from CoGeNT~\cite{Aalseth:2010vx}, which, if interpreted as
dark matter, also favor the same low mass $m_X \sim 5-10~\gev$ and
high cross section $\sigmaSI \sim 10^{-4}~\pb$ region of dark matter
parameter space.  These results are consistent with recent bounds and
results from CDMS~\cite{Ahmed:2009zw}.  The consistency of several
direct detection experiments is, of course, important to establish a
dark matter signal.  Improved statistics will be essential, but given
the difficulty of making definitive background determinations,
independent confirmation by completely different means is also highly
desirable.  As discussed above, the direct detection data may
naturally be explained by scalar dark matter interacting with exotic
4th generation quarks. Data already taken by the Super-Kamiokande
experiment provide a promising probe of these interpretations of DAMA,
CDMS, and CoGeNT through indirect detection~\cite{SuperKlimit}.  The
results derived here show that these explanations, including WIMPless
models, can also be tested very directly with data already taken at
the Tevatron and have extraordinarily promising implications for
early runs of the LHC.

\section*{Acknowledgments}

We gratefully acknowledge H.~Baer, A.~Barr, G.~Kribs, B.~Nelson,
T.~Tait, X.~Tata, D.~Toback and D.~Whiteson for useful discussions.
JK is grateful to George Mitchell, KIAS, and the KEK Theory Center for
their hospitality while this work was in preparation.  The work of JA
was supported by NCTS, grant number NSC 98--2119--M--002--001. The
work of JLF was supported in part by NSF grant PHY--0653656.  The work
of SS was supported in part by the Department of Energy under
Grant~DE-FG02-04ER-41298.

\appendix

\section*{Appendix: Impact of Cuts on Signal and Backgrounds}

In this Appendix, we present tables listing the cross sections after
cuts for the $T' \bar{T}'$ signal and the main SM backgrounds.  In the
upper section of each table, each line gives the cross section after
including all cuts above.  In the lower section, each line gives the
cross section after including the cut on that line, and all precuts.
For the signal, three examples with $m_X = 1~\gev$ and $m_{T'} = 300$,
400, and 500 GeV are chosen.  The $W$ and $Z$ cross sections in
parentheses were simulated with a cut on $\met > 80~\gev$ and at least
3 jets in the parton-level generation.

For the LHC, we have divided the $t\bar t$ background according to
decays: hadronic, single tau lepton, semi-leptonic (electron or muon),
double leptonic (electron or muon), and double tau lepton. Only the
contributing decay modes have been included in the table.

\begin{table}[htbp]
\caption{Signal and background cross sections in fb after cuts for the
semi-leptonic channel at the Tevatron.  The signal examples are for
$m_X = 1~\gev$ and $m_{T'} = 300$, 400, and 500 GeV as indicated.  The
$W$ and $Z$ cross sections in parentheses were simulated with a cut on
$\met > 80~\gev$ and at least 3 jets in the parton-level generation.
}
\begin{tabular}{|c|c|c|c|c|c|c|}
\hline
Cut & $T'$ (300) & $T'$ (400) & $T'$ (500) & $t\bar t$ & $W$+jets & $Z$+jets \\
\hline
No cut         & 203.2 & 16.33 & 1.11 & 5619 & (5230) & (132) \\
1 $\mu$/$e$, no $\tau$ & 36.1 & 2.88 & 0.194 & 1041 & (2062) & (15.7) \\
$\met > 100~\gev$ & 17.7 & 2.00 & 0.157 & 107.2 & (730.2) & (3.7) \\
$m_T^W > 100~\gev$ & 10.7 & 1.38 & 0.114 & 22.6 & (36.8) & - \\
$\ge 4$ jets & 4.81 & 0.64 & 0.062 & 2.6 & 0.29 & - \\
$|m_{jj}-m_W| < 10~\gev$ & 4.13 & 0.51 & 0.049 & 2.2 & 0.19 & - \\
\hline
All precuts & 4.13 & 0.51 & 0.049 & 2.19 & 0.19 & - \\
$m_T^W > 150~\gev$ & 1.93 & 0.325 & 0.036 & 0.62 & 0.035 & - \\
$\met > 150~\gev$ & 1.75 & 0.367 & 0.041 & 0.281 & 0.035 & - \\
$H_T > 300~\gev$ & 1.93 & 0.353 & 0.042 & 1.18 & 0.07 & - \\
$\met > 150$, $H_T > 300$ & 1.04 & 0.279 & 0.037 & 0.056 & 0.017 & - \\
\hline
\end{tabular}
\label{table:TeVsemi}
\end{table}

\begin{table}[htbp]
\caption{As in \tableref{TeVsemi}, but for the semi-leptonic channel
at the 10 TeV LHC and with cross sections in pb.}
\begin{tabular}{|c|c|c|c|c|c|c|c|}
\hline
Cut & $T'$ (300) & $T'$ (400) & $T'$ (500) & $t\bar t$
(1 $e/\mu$) & $t\bar t$ (1 $\tau$) & $t\bar t$ (2 $e/\mu$) & $W$+jets \\
\hline
No cut & 14.89 & 3.16 & 0.922 & 66.67 & 43.96 & 10.62 & (42.28) \\
1 $\mu$/$e$, no $\tau$ & 3.2 & 0.669 & 0.193 & 36.45 & 8.15 & 3.18 & (15.74) \\
$\met > 100~\gev$ & 1.92 & 0.52 & 0.165 & 5.05 & 2.07 & 0.888 & (10.33) \\
$m_T^W > 100~\gev$ & 1.1 & 0.342 & 0.116 & 0.134 & 0.638 & 0.471 & (0.235) \\
$\ge 4$ jets & 0.357 & 0.116 & 0.043 & 0.056 & 0.091 & 0.062 & 0.028 \\
$|m_{jj}-m_W| < 10~\gev$ & 0.165 & 0.049 & 0.016 & 0.026 & 0.03 & 0.014 & 0.01 \\
\hline
All precuts & 0.165 & 0.049 & 0.016 & 0.027 & 0.031 & 0.014 & 0.01 \\
$m_T^W > 150~\gev$ & 0.081 & 0.033 & 0.012 & 0.001 & 0.015 & 0.006 & 0.002 \\
$m_T^W > 200~\gev$ & 0.032 & 0.019 & 0.008 & 0 & 0.006 & 0.003 & 0.001 \\
$\met > 150~\gev$ & 0.099 & 0.036 & 0.014 & 0.005 & 0.012 & 0.005 & 0.004 \\
$\met > 200~\gev$ & 0.041 & 0.025 & 0.01 & 0.001 & 0.004 & 0.001 & 0.002 \\
$\met > 250~\gev$ & 0.016 & 0.013 & 0.007 & 0 & 0.002 & 0.001 & 0.002 \\
$H_T > 400~\gev$ & 0.107 & 0.035 & 0.013 & 0.018 & 0.022 & 0.008 & 0.007 \\
$H_T > 500~\gev$ & 0.059 & 0.021 & 0.009 & 0.011 & 0.014 & 0.004 & 0.005 \\
$\met > 150$, $H_T > 400$ & 0.067 & 0.027 & 0.012 & 0.004 & 0.01 & 0.004 & 0.003 \\
$\met > 150$, $H_T > 500$ & 0.037 & 0.016 & 0.008 & 0.003 & 0.007 & 0.002 & 0.002 \\
$\met > 200$, $H_T > 400$ & 0.032 & 0.02 & 0.009 & 0.001 & 0.004 & 0.001 & 0.002 \\
$\met > 200$, $H_T > 500$ & 0.02 & 0.013 & 0.007 & 0 & 0.004 & 0.001 & 0.002 \\
$\met > 250$, $H_T > 400$ & 0.014 & 0.012 & 0.006 & 0 & 0.002 & 0.001 & 0.002 \\
$\met > 250$, $H_T > 500$ & 0.009 & 0.008 & 0.005 & 0 & 0.002 & 0.001 & 0.001 \\
\hline
\end{tabular}
\label{table:LHCsemi}
\end{table}

\begin{table}[htbp]
\caption{As in \tableref{TeVsemi}, but for the hadronic channel at the
Tevatron and with cross sections in fb.}
\begin{tabular}{|c|c|c|c|c|c|c|}
\hline
Cut & $T'$ (300) & $T'$ (400) & $T'$ (500) & $t\bar t$ & $W$+jets & $Z$+jets \\
\hline
No cut & 203.24 & 16.33 & 1.11 & 5619.1 & (5179.06) & (3030.09) \\
0 isolated leptons & 82.88 & 6.97 & 0.499 & 2265.54 & (1756.96) & (2545.12) \\
$\met > 100~\gev$ & 42.86 & 5.28 & 0.422 & 125.93 & (663.5) & (1219.22) \\
$\ge 5$ jets & 22.64 & 3.07 & 0.273 & 22.11 & 3.3 & 2.6 \\
$\Delta\phi$ cuts & 19.0 & 2.74 & 0.245 & 15.8 & 2.8 & 2.2 \\
\hline
All precuts & 19 & 2.74 & 0.245 & 15.8 & 2.8 & 2.2 \\
$\met > 150~\gev$ & 7.93 & 2.04 & 0.21 & 4.32 & 0.791 & 0.93 \\
$\met > 200~\gev$ & 1.06 & 1.25 & 0.158 & 1.02 & 0.183 & 0.313 \\
$\met > 250~\gev$ & 0.142 & 0.516 & 0.109 & 0.347 & 0.025 & 0.162 \\
$H_T > 300~\gev$ & 9.9 & 2.04 & 0.224 & 5.16 & 0.55 & 0.495 \\
$H_T > 350~\gev$ & 4.92 & 1.37 & 0.182 & 2.72 & 0.208 & 0.162 \\
$H_T > 400~\gev$ & 2.46 & 0.787 & 0.135 & 1.22 & 0.083 & 0.081 \\
$\met > 150$, $H_T > 300$ & 5.2 & 1.64 & 0.197 & 2.19 & 0.217 & 0.404 \\
$\met > 200$, $H_T > 300$ & 0.996 & 1.11 & 0.153 & 0.821 & 0.067 & 0.212 \\
$\met > 250$, $H_T > 300$ & 0.142 & 0.495 & 0.108 & 0.347 & 0.025 & 0.142 \\
$\met > 200$, $H_T > 350$ & 0.711 & 0.794 & 0.131 & 0.511 & 0.033 & 0.081 \\
$\met > 250$, $H_T > 350$ & 0.142 & 0.399 & 0.098 & 0.255 & 0.017 & 0.071 \\
$N$(jets) $\ge 6$ & 8.45 & 1.3 & 0.125 & 3.1 & 0.333 & 0.212 \\
$N$(jets) $\ge 6$, $\met > 150~\gev$ & 3.62 & 0.957 & 0.107 & 0.948 & 0.092 & 0.101 \\
$N$(jets) $\ge 6$, $\met > 200~\gev$ & 0.467 & 0.583 & 0.08 & 0.237 & 0.025 & 0.04 \\
$N$(jets) $\ge 6$, $H_T > 300~\gev$ & 4.84 & 0.995 & 0.116 & 1.28 & 0.092 & 0.081 \\
$N$(jets) $\ge 6$, $H_T > 350~\gev$ & 2.34 & 0.683 & 0.097 & 0.693 & 0.05 & 0.02 \\
$N$(jets) $\ge 6$, $H_T > 400~\gev$ & 1.16 & 0.364 & 0.072 & 0.328 & 0.017 & 0.01 \\
$N(j),\met, H_T > 6, 150, 300$ & 2.64 & 0.786 & 0.102 & 0.58 & 0.033 & 0.061 \\
\hline
\end{tabular}
\label{table:TeVhad}
\end{table}

\begin{table}[htbp]
\caption{As in \tableref{TeVsemi}, but for the hadronic channel at
the 10 TeV LHC and with cross sections in pb.}
\begin{tabular}{|c|c|c|c|c|c|c|c|c|}
\hline
Cut & $T'$ (300) & $T'$ (400) & $T'$ (500) &
$t\bar t$ (1 $\tau$) & $t\bar t$ (1 $e/\mu$) & $t\bar t$ (had) & $W$+jets & $Z$+jets \\
\hline
No cut & 14.89 & 3.16 & 0.922 & 43.96 & 66.67 & 104.59 & (42.28) & (18.86) \\
0 isolated leptons & 6.75 & 1.5 & 0.45 & 16.88 & 13.11 & 72.29 & (16.8) & (15.71) \\
$\met > 100~\gev$ & 4.15 & 1.21 & 0.394 & 3.91 & 2.67 & 0.097 & (11.25) & (11.48) \\
$\ge 5$ jets & 1.34 & 0.406 & 0.135 & 0.664 & 0.47 & 0.031 & 0.305 & 0.212 \\
$\Delta\phi$ cuts & 1.19 & 0.374 & 0.125 & 0.56 & 0.41 & 0.01 & 0.265 & 0.187 \\
\hline
All precuts & 1.19 & 0.374 & 0.125 & 0.56 & 0.41 & 0.01 & 0.265 & 0.187 \\
$\met > 150~\gev$ & 0.727 & 0.341 & 0.136 & 0.205 & 0.128 & - & 0.131 & 0.119\\
$\met > 200~\gev$ & 0.291 & 0.231 & 0.107 & 0.069 & 0.042 & - & 0.06 & 0.069\\
$\met > 250~\gev$ & 0.107 & 0.131 & 0.079 & 0.026 & 0.015 & - & 0.026 & 0.04\\
$\met > 300~\gev$ & 0.043 & 0.062 & 0.053 & 0.011 & 0.005 & - & 0.014 & 0.022\\
$H_T > 400~\gev$ & 1.02 & 0.379 & 0.149 & 0.422 & 0.307 & - & 0.207 & 0.145\\
$H_T > 500~\gev$ & 0.668 & 0.264 & 0.118 & 0.275 & 0.209 & - & 0.133 & 0.096\\
$\met > 150$, $H_T > 400$ & 0.6 & 0.301 & 0.128 & 0.176 & 0.109 & - & 0.113 & 0.1\\
$\met > 150$, $H_T > 500$ & 0.411 & 0.213 & 0.103 & 0.129 & 0.082 & - & 0.078 & 0.071\\
$\met > 200$, $H_T > 400$ & 0.271 & 0.21 & 0.103 & 0.065 & 0.039 & - & 0.056 & 0.062\\
$\met > 200$, $H_T > 500$ & 0.213 & 0.152 & 0.085 & 0.053 & 0.03 & - & 0.042 & 0.049\\
$\met > 250$, $H_T > 400$ & 0.106 & 0.126 & 0.078 & 0.026 & 0.015 & - & 0.025 & 0.038\\
$\met > 250$, $H_T > 500$ & 0.097 & 0.096 & 0.067 & 0.024 & 0.012 & - & 0.021 & 0.031\\
$\met > 300$, $H_T > 400$ & 0.043 & 0.06 & 0.053 & 0.011 & 0.005 & - & 0.014 & 0.021\\
$\met > 300$, $H_T > 500$ & 0.043 & 0.05 & 0.048 & 0.011 & 0.005 & - & 0.013 & 0.019\\
$N$(jets) $\ge 6$ & 0.509 & 0.181 & 0.064 & 0.178 & 0.13 & - & 0.046 & 0.028\\
$N$(jets) $\ge 6$, $\met > 150$  & 0.278 & 0.138 & 0.055 & 0.068 & 0.044 & - & 0.027 & 0.019\\
$N$(jets) $\ge 6$, $\met > 200$  & 0.134 & 0.096 & 0.044 & 0.025 & 0.015 & - & 0.015 & 0.012\\
$N$(jets) $\ge 6$, $\met > 250$  & 0.052 & 0.055 & 0.034 & 0.01 & 0.005 & - & 0.006 & 0.008\\
$N$(jets) $\ge 6$, $\met > 300$  & 0.023 & 0.027 & 0.024 & 0.004 & 0.001 & - & 0.003 & 0.005\\
$N$(jets) $\ge 6$, $H_T > 400$  & 0.424 & 0.166 & 0.062 & 0.152 & 0.109 & - & 0.041 & 0.025\\
$N$(jets) $\ge 6$, $H_T > 500$  & 0.319 & 0.126 & 0.052 & 0.109 & 0.08 & - & 0.03 & 0.019\\
$N(j),\met,H_T > 6,150,400$ & 0.25 & 0.128 & 0.053 & 0.063 & 0.04 & - & 0.025 & 0.018\\
$N(j),\met,H_T > 6,150,500$ & 0.202 & 0.099 & 0.045 & 0.049 & 0.03 & - & 0.019 & 0.014\\
$N(j),\met,H_T > 6,200,400$ & 0.126 & 0.09 & 0.043 & 0.024 & 0.014 & - & 0.014 & 0.012\\
$N(j),\met,H_T > 6,200,500$ & 0.115 & 0.069 & 0.038 & 0.021 & 0.011 & - & 0.011 & 0.01\\
$N(j),\met,H_T > 6,250,400$ & 0.051 & 0.054 & 0.033 & 0.01 & 0.005 & - & 0.006 & 0.007\\
$N(j),\met,H_T > 6,250,500$ & 0.048 & 0.044 & 0.03 & 0.009 & 0.004 & - & 0.005 & 0.007\\
$N(j),\met,H_T > 6,300,400$ & 0.023 & 0.027 & 0.023 & 0.004 & 0.001 & - & 0.003 & 0.005\\
$N(j),\met,H_T > 6,300,500$ & 0.023 & 0.023 & 0.022 & 0.004 & 0.001 & - & 0.003 & 0.005\\
\hline
\end{tabular}
\label{table:LHChad}
\end{table}

\clearpage



\end{document}